\documentclass[fleqn,usenatbib]{mnras}
\usepackage[T1]{fontenc}
\usepackage{graphicx}
\usepackage{amsmath}
\usepackage{newtxtext,newtxmath}   
\usepackage[mathscr]{eucal}         
\usepackage{bm}
\usepackage{tensor}
\usepackage{booktabs}
\usepackage{array}                  
\usepackage{adjustbox}              
\usepackage{xcolor}
\usepackage{enumitem}
\hypersetup{linkcolor=magenta, citecolor=blue, urlcolor=blue}

\def\equationautorefname~#1\null{equation~(#1)\null}

\title[Amplitude-growth degeneracy in background-inert $f(Q)$]{The Amplitude-Growth Degeneracy and Implied \texorpdfstring{$A_s$}{Lg} Diagnostic for Background-Inert Modified Gravity}
\author[A. Kolhatkar and P. K. Sahoo]{
Ameya Kolhatkar,$^{1}$
P. K. Sahoo$^{1}$\thanks{E-mail: pksahoo@hyderabad.bits-pilani.ac.in}
\\
$^{1}$Department of Mathematics, Birla Institute of Technology and Science, Pilani,\\ Hyderabad Campus, Jawahar Nagar, Kapra Mandal, Medchal District, Telangana 500078, India
}
\date{Accepted XXX. Received YYY; in original form ZZZ}
\pubyear{2026}

\begin{document}
\label{firstpage}
\pagerange{\pageref{firstpage}--\pageref{lastpage}}
\maketitle

\begin{abstract}
    We \textbf{show} that any background-inert perturbative coupling $ \lambda $ in coincident $ f(Q) $ gravity exhibits an approximate degeneracy with the clustering amplitude $ \sigma_{80} $, when using compressed CMB distance priors. This follows from the invariance of $\sigma_{80}$ under the simultaneous rescaling of the primordial amplitude $A_s$ and the growth factor $D_0(\lambda)$, which is partially broken by redshift dependent probes like $f\sigma_8$ and $E_g$. We outline a consistency check scheme, whose logic extends to any model in which the coupling is background inert, by computing $ A_s $ per posterior sample, needed to reproduce the $ \sigma_{80} $ preferred by the sampler. We perform our analysis with two dataset pipelines, based on the coupled/decoupled $ f\sigma_8(z) $ data. To ensure theoretical diversity, we include $ \Lambda $CDM and the Hybrid model in the $ f(Q) $ framework. Our results illustrate that adding the $ \lambda_0\sqrt{QQ_0} $ correction to the models inflates $ \sigma_{80} $ by $ 5\%-8\% $ as compared to its vanilla variant, while the Bayesian evidence disfavors every alternative considered - $\Delta\log\mathcal{Z} = -0.5$ to $-2.6$ against $\Lambda$CDM on the same pipeline. Propagating this inflated $ \sigma_{80} $ through a per-sample computation of the implied primordial amplitude accounting for both the transfer function and the modified growth factor yields $\ln(10^{10}A_s)$ in $1.5\sigma-2.6\sigma$ tension with Planck 2018. Imposing the $ \ln(10^{10}A_s) $ constraint from Planck 2018 as an additional prior removes this inflation, pulling $ \sigma_{80} $ to the Planck value and $ \lambda_0 $ to values consistent with $ 0 $, with the implied amplitude recovering to within $ 0.2\sigma $ of Planck in every case. We find no model-dataset combination preferred over $ \Lambda $CDM.
\end{abstract}


\section{Introduction}\label{sec_Intro}
The early universe has been pinned down by $ \Lambda $CDM cosmology through an accurate description of Big Bang Nucleosynthesis (BBN) \cite{Cyburt:2015mya} and the acoustic peaks of the Cosmic Microwave Background (CMB) \cite{Planck:2018vyg}. In addition, recent large Scale Structure (LSS) studies \cite{eBOSS:2020yzd} show excellent agreement of the concordance model with the data. However, as precision cosmological probes are advancing rapidly, statistically significant discrepancies have been bolstered with newer surveys. Alongside the well known $ H_0 $ tension between the early time CMB measurement and the local measurement by SH0ES \cite{Riess:2019cxk}, the $ S_8=\sigma_{80}\sqrt{\Omega_{m0}/0.3} $ tension has become a central problem in modern cosmology. Late-time probes such as Weak Lensing (WL) \cite{Heymans:2020gsg} and Redshift-Space Distortions (RSD) \cite{eBOSS:2020yzd} prefer a lower clustering amplitude than the CMB extrapolation, suggesting a systematic suppression of the late-time growth of structure compared to $ \Lambda $CDM predictions. Choice of observational inputs also affects the evidence for a non-$\Lambda$CDM cosmology. With DESI DR2 BAO, the preference for dynamical dark energy varies from below $3.0$ to about $3.8\sigma$ depending on the Supernova sample \citep{Capozziello:2025qmh}.

These and other issues with $ \Lambda $CDM have motivated the investigation of modified theories of gravity. Among these, Symmetric Teleparallel Gravity and its extension, $ f(Q) $ gravity \citep{Nester:1998mp, BeltranJimenez:2017tkd, BeltranJimenez:2019esp, BeltranJimenez:2019tme} have garnered significant attention \citep{BeltranJimenez:2018vdo, Lu:2019hra, Mandal:2020buf, Mandal:2020lyq, Barros:2020bgg, Lin:2021uqa, Dimakis:2021gby, Hassan:2021egb, Atayde:2021pgb, Frusciante:2021sio, Hassan:2022hcb, Sokoliuk:2023ccw, Gadbail:2023loj, Atayde:2023aoj, Shabani:2023xfn, Mishra:2024rut, Capozziello:2024vix, Capozziello:2024jir, Mishra:2024shg, Boiza:2025xpn, Kolhatkar:2025ubm, Capozziello:2025hyw, Kavya:2025vsj, Li:2025msm, Kolhatkar:2026ixl, Kolhatkar:2026bss}. By attributing gravity entirely to the non-metricity scalar $ Q $ for a flat, torsion free connection, $ f(Q) $ theories offer a phenomenological framework to explain the late-time dynamics without invoking any exotic fields. The functional form $f(Q)$ can be reconstructed from the expansion history preferred by data through a technique called cosmography \citep{Capozziello:2022wgl}. Such a process relies on $H(z)$ and its derivatives to reconstruct the Lagrangian and is thus blind to the background-inert square-root term explored in this work. 

The perturbative viability of cosmology in $ f(Q) $ theory is contested at the level of fundamental theory. It has been shown \citep{Gomes:2023tur} (see also \cite{Heisenberg:2023wgk, BeltranJimenez:2026wja}) that all the branches of the theory either have ghosts or strong coupling issues in the spatially flat FLRW case. A ghost-free realization of the Symmetric Teleparallel cosmology is constructed in \cite{Bello-Morales:2024vqk}, which is however, built on the quadratic object $Q_\alpha Q^\alpha$ invariant under transverse diffeomorphisms and therefore lies outside the $f(Q)$ family considered here. Within $f(Q)$ itself, the linear branch $f_{QQ}=0$ remains the only case free of the pathologies identified in \cite{Gomes:2023tur}. We thus treat the $ \sqrt{QQ_0} $ term as a phenomenological realization of a background inert $ \mu_G(z) $, and quantify in \autoref{subsec_EFTandQSA} the proximity of the said models to the $ f_{QQ} = 0 $ boundary. 

The standard methodology for assessing physical viability involves the use of Bayesian inference using late-time probes, typically consisting of Type Ia Supernovae, BAO and RSD. As shown in \cite{Zhai:2019nad}, the compressed CMB likelihood suffices as an alternative to the full CMB implementation for probing late-time dynamics, putting tight constraints on the shift parameters $ R $ and $ l_a $ along with the baryon energy density $ \Omega_{b0}h^2 $ and the spectral index $ n_s $. However, this methodological ``shortcut" introduces a critical vulnerability. The shift parameters only constrain the background geometry and carry no information about the primordial amplitude $A_s$, which is left free, thereby blinding the pipeline to the early Universe. 

The present work aims to demonstrate that this phenomenon arises as a statistical degeneracy between the amplitude and the growth factor, if the theory converges to standard $ \Lambda $CDM before recombination and has a perturbative coupling inert to the background. This coupling forms a degeneracy with the clustering amplitude, along which the sampler can raise $ \sigma_{80} $ at a negligible cost to the fit. Specifically, we analyze the $ \lambda_0\sqrt{QQ_0} $ correction to the $ f(Q) $ Lagrangian covering both $ \Lambda $CDM and the Hybrid model, which is a boundary term invisible to the background $ H(z) $. Out of many studies in the past, two recent studies \cite{Li:2025msm, Kolhatkar:2026bss} in particular investigate the effects of the square-root correction under growth data, although without the compressed CMB likelihood. The results in both the manuscripts show a mild form of the $ \sigma_{80} $ inflation, but neither of them actually explain its origin. The present work identifies the cause of the degeneracy and provides the diagnostic that breaks it. 

We propose the ``implied $ A_s $" diagnostic -- a consistency check that maps the late-time clustering amplitude preferred by the sampler back to the primordial amplitude it requires. Because this mapping depends on both the transfer function and the modified growth factor $ D_0(\lambda) $, it must be evaluated per posterior sample rather than through a universal rescaling. Applied to both models with the square-root correction, we find that including $ \lambda_0 $ inflates $ \sigma_{80} $ by $ \sim 5\%-8\% $ above its plain variant baseline while leaving the Information Criterion mutually inconsistent - $ \Delta $AIC and $ \Delta $DIC show weak to no preference over $ \Lambda $CDM while $ \Delta $BIC and $ \Delta\log(\mathcal{Z}) $ favors $ \Lambda $CDM in the moderate to strong range. The proposed diagnostic resolves this ambiguity by showing that the inflated amplitude requires the corresponding $ \ln(10^{10}A_s) $ value to be in $ 1.5\sigma-2.6\sigma $ tension with Planck 2018 results. Imposing the same Planck constraint on $ \ln(10^{10}A_s) $ as a per-sample prior then breaks the degeneracy - the $ \lambda_0 $ interval contracts by a factor of $ 3 $ while becoming consistent with zero, and the implied $ A_s $ returns to within $ 0.2\sigma $ of Planck with no preference over $ \Lambda $CDM. We therefore recommend the amplitude prior as a routine safeguard for compressed CMB analyses of background-inert couplings, where the degeneracy would otherwise leave the model-selection verdict undetermined.

This paper is organized as follows. \autoref{sec_Theroetical_Framework} presents the theoretical framework, including coincident $ f(Q) $ cosmology, the significance of the $ \sqrt{Q} $ term, the models employed, and the implied $ A_s $ diagnostic. \autoref{sec_DM} describes the datasets used and the statistical methodology. \autoref{sec_results} presents the posterior contours, parameter constraints and evidence metric values along with the demonstration of the $ \lambda_0 $ induced degeneracy. \autoref{sec_discussion} contains an extensive discussion on the results obtained. Finally, \autoref{sec_conclusion} concludes the manuscript with closing remarks.
\section{Theoretical Framework}\label{sec_Theroetical_Framework}

\subsection{Coincident \texorpdfstring{$ f(Q) $}{Lg} Cosmology and the \texorpdfstring{$ \sqrt{Q} $}{Lg} term}\label{subsec_fQ}
We begin by writing down the action of $ f(Q) $ theory
\begin{equation}\label{eq_fQ_action}
    S = \int d^4x\sqrt{-g}\left( \frac{f(Q)}{2\kappa} + \mathcal{L_M} \right)\;,
\end{equation}
where $ \kappa = 8\pi G $, $ \mathcal{L_M} $ is the Lagrangian of the matter fields and $ f(Q) $ is an arbitrary function constructed out of the non-metricity scalar $ Q = Q_{\alpha\beta\gamma}P^{\alpha\beta\gamma} $. The superpotential $ \tensor{P}{^\alpha_{\mu\nu}} $ is defined in terms of the non-metricity tensor $ Q_{\alpha\beta\gamma}=\nabla_\alpha g_{\beta\gamma} $ as
\begin{equation}\label{eq_superpotential}
    \tensor{P}{^\alpha_{\mu\nu}} = -\frac{1}{2}\tensor{L}{^\alpha_{\mu\nu}} + \frac{1}{4} g_{\mu\nu} ( Q^\alpha - \Tilde{Q}^\alpha ) - \frac{1}{4} \tensor{\delta}{^\alpha_{(\mu}} \tensor{Q}{_{\nu)}}\;.
\end{equation}
The two independent traces of $ Q_{\alpha\beta\gamma} $ are $ Q_\alpha = g^{\mu\nu}Q_{\alpha\mu\nu} $ and $ \Tilde{Q}_\alpha = g^{\mu\nu}Q_{\mu\nu\alpha} $. Varying \autoref{eq_fQ_action} with respect to the inverse metric yields the general field equations
\begin{multline}\label{eq_fQfieldequations}
    \frac{2}{\sqrt{-g}}\nabla_\alpha(\sqrt{-g}f_Q\tensor{P}{^\alpha_{\mu\nu}})-\frac{1}{2}g_{\mu\nu}f+\\
f_Q(P_{\mu\alpha\beta}\tensor{Q}{_\nu^{\alpha\beta}}-2Q_{\alpha\beta\mu}\tensor{P}{^{\alpha\beta}_\nu})=\kappa T_{\mu\nu}\;.
\end{multline}
For our cosmological analysis, we chose a spatially flat FLRW metric described by the line element
\begin{equation}\label{eq_FLRW}
    ds^2 = -dt^2 + a^2(t) (dx^2 + dy^2 + dz^2)\;.
\end{equation}
In this Universe expanding with the scale factor $ a(t) $, \autoref{eq_fQfieldequations} becomes the modified Friedmann equations 
\begin{gather}
    3H^2 = \kappa (\rho + \rho_D)\;,\label{eq_modified_friedmann1}\\
    3H^2 + 2\dot{H} = -\kappa(p+p_D)\;,\label{eq_modified_friedmann2}
\end{gather}
with the energy density and pressure associated with the $ f(Q) $ geometry is expressed as 
\begin{gather}
    2\kappa\rho_D = Q(1-2f_Q) + f\;,\label{eq_fQ_rho}\\
    2\kappa p_D = 4\dot{H}(f_Q - 1) - f + Q(8f_{QQ}\dot{H} + 2f_Q - 1)\label{eq_fQ_p}\;.
\end{gather}
Taking a closer look at \autoref{eq_fQ_rho}, one finds that $ \rho_D $ and $ p_D $ vanish for $ f(Q) \propto \sqrt{Q} $. This is expected since, for a spatially flat FLRW metric, $ Q = 6H^2 $ and $ \sqrt{-g}\sqrt{Q} = a^2 \dot{a} = \frac{1}{3}\frac{d}{dt} (a^3) $ being a total derivative contributes nothing to the background expansion. However, this only holds for a perfectly isotropic and homogeneous spacetime, and once perturbations are introduced, the $ \sqrt{Q} $ term modifies the growth of structure, albeit in complete isolation from background expansion. This modification is captured by the effective gravitational constant $ G_{eff} = \mu_G G_N = G_N / f_Q $\footnote{We are inserting the subscript `G' in order to differentiate the relative strength ratio of gravity from the distance modulus in \autoref{subsec_datasets}}. 

\subsection{Models used}\label{subsec_models}
In this work, we employ two models. Throughout this manuscript, we use $ \Omega(z) = \Omega_{m0}(1+z)^3 + \Omega_{r0}(1+z)^4 $ and $ \Tilde{\Omega}(z) = \Omega(z) + \alpha_2 $ and their present day values are given by $ \Omega_0 = \Omega_{m0} + \Omega_{r0} $ and $ \Tilde{\Omega}_0 = \Omega_0 + \alpha_2 $. 
\begin{enumerate}
    \item $ \bm{\Lambda} $\textbf{CDM} : We use this as a reference for all our results. The equivalent $ f(Q) $ Lagrangian is $ Q + (1-\Omega_0) Q_0 $ and the Hubble parameter is 
    \begin{equation}\label{eq_hubble_lcdm}
        H(z) = H_0 \sqrt{\Omega(z) + 1 - \Omega_0}\;.
    \end{equation}
    \item $ \bm{f(Q)} $ : The Hybrid model Lagrangian reads $ Q + \alpha_2 Q_0 + \frac{1}{3}(1-\Tilde{\Omega}_0) \frac{Q_0^2}{Q} $. \cite{Kolhatkar:2026bss} explicitly proves why the linear coupling must be exactly equal to unity to avoid early time phantom deviations. Moreover, the indistinguishable background of the Hybrid $ f(Q) $ model from $ \Lambda $CDM was also independently confirmed by MCMC analysis in the aforementioned study. The Hubble parameter is 
    \begin{equation}\label{eq_hubble_hybrid}
        H(z) = H_0 \sqrt{\frac{\Tilde{\Omega}(z) + \sqrt{\Tilde{\Omega}^2(z) + 4(1-\Tilde{\Omega}_0)}}{2}}\;.
    \end{equation}\
\end{enumerate}
For each of the two models, we employ three variants
\begin{enumerate}
    \item \textbf{Model}: Plain model 
    \item \textbf{Model} $ \bm{+ \lambda_0} $: With the $ \lambda_0\sqrt{QQ_0} $ correction to the Lagrangian 
    \item \textbf{Model} $ \bm{+ \lambda_0 + \ln(A_s)} $: With Planck priors on top of the square-root correction.
\end{enumerate}
This gives six combinations, which we analyze in \autoref{sec_results}. Expressions for the effective gravitational coupling for both models are as follows
\begin{gather}
    \mu_{G,\;\Lambda\text{CDM}}^{-1}(z,\lambda_0) = 1 + \frac{\lambda_0H_0}{2H(z)}\;,\label{eq_muG_lcdm}\\
    \mu_{G,\;f(Q)}^{-1}(z, \lambda_0) = 1 - (1-\Tilde{\Omega}_0)\frac{H_0^4}{3H^4(z)} + \frac{\lambda_0H_0}{2H(z)}\;.\label{eq_muG_fQ}
\end{gather}
\subsection{Strong coupling, EFT diagnostics and Quasi-Static Approximation validity}\label{subsec_EFTandQSA}
We define the Pathology proximity parameter that measures the proximity of our models to a healthy, ghost-free boundary: 
\begin{equation}\label{eq_epsilonsc}
    \varepsilon_{sc}(z) \equiv \frac{f_{QQ}(z)Q(z)}{f_Q(z)}\;.
\end{equation}
This is the structural analog of the Song-Hu-Sawicki B-parameter of $ f(R) $ given by $ B = (f_{RR}/f_R) \cdot R'H/H' $ in \cite{Song:2006ej}, and in the $ f(Q) $ case, $ B_{f(Q)} = (f_{QQ}/f_Q) \cdot Q'H/H' = 2\varepsilon_{sc} $. The primary feature of it is that for a Lagrangian linear in $ Q $, $ f_{QQ} = 0\implies \varepsilon_{sc} = 0 $, agreeing with the result from \cite{Gomes:2023tur} that such a class of models do not exhibit a strong coupling issue. For non-linear models, $ |\varepsilon_{sc}|\ll1 $ indicates closeness to the ``healthy" ghost-free/strong coupling free boundary $ \varepsilon_{sc}=0 $. We emphasize that this is a  heuristic argument and does not serve as a proxy for full EFT analysis. The closed-form expressions for the models are
\begin{gather}
    \varepsilon^{-1}_{sc,\;\Lambda\text{CDM}}(z) = -2-\frac{4H(z)}{\lambda_0 H_0}\label{eq_epssc_lcdm}\;,\\
    \varepsilon^{-1}_{sc,f(Q)}(z) = \frac{1-\frac{1}{3}(1-\Tilde{\Omega}_0)\frac{H_0^4}{H^4(z)} + \frac{\lambda_0 H_0}{2H(z)}}{\frac{2}{3}(1-\Tilde{\Omega}_0)\frac{H_0^4}{H^4(z)} - \frac{\lambda_0 H_0}{4H(z)}} \label{eq_epssc_fQ}\;.
\end{gather}
This quantity $ |\varepsilon_{sc}| $ is monotonic for the $ \Lambda $CDM variant class, while for the Hybrid class of models, we numerically verify over $ z\in[0,3] $ that the maximum coincides with $ z=0 $ within $ 1.5\% $ across the posterior. Thus, the extremal parameter values can be computed with the exact values of model parameters obtained from Bayesian analysis further down this work using
\begin{gather}
    \varepsilon^{-1}_{sc0,\;\Lambda\text{CDM}} = -2-\frac{4}{\lambda_0}\label{eq_epssc0_lcdm}\;,\\
    \varepsilon^{-1}_{sc0,f(Q)} = \frac{1-\frac{1}{3}(1-\Tilde{\Omega}_0) + \frac{\lambda_0}{2}}{\frac{2}{3}(1-\Tilde{\Omega}_0) - \frac{\lambda_0}{4}}\label{eq_epssc0_fQ}\;.
\end{gather}
The perturbative observables, $ f\sigma_8(z) $ and $ E_g(z) $ are computed from the scale-independent coupling $ \mu_G(z) = 1/f_Q $ as described in the previous section. This follows after taking the sub-horizon quasi-static limit of the full perturbation equations as in \cite{BeltranJimenez:2019tme}, which also mentions that the quasi-static limit is not well defined in the present theories due to strong coupling reasons discussed above. We thus treat this as a working approximation and estimate the error it incurs, rather than assert its validity. Taking the quasi-static approximation (QSA) retains the $ k^2 $ dependent terms suppressing the discarded terms by a factor of $ (aH/ck)^2 $. Over the redshift range spanned by the growth data this ratio is maximized at $ z=0 $, where it reduces to $ (H_0/ck)^2 $ independently of cosmology, giving $1.1\times10^{-5}$ at $ k=0.1 h\,\mathrm{Mpc}^{-1}$ and $1.1\times10^{-3}$ at $ k=0.01 h\,\mathrm{Mpc}^{-1}$. Extending to the Lyman-$\alpha$ BAO reach raises it by at most $4\%$, the Hybrid background being the largest. The ratio reaches $1\%$ only below $k\simeq4\times10^{-3}h\,\mathrm{Mpc}^{-1}$, a mode of $\sim1600\,h^{-1}\mathrm{Mpc}$, larger than any scale probed here.


\subsection{The implied \texorpdfstring{$ A_s $}{Lg} consistency criterion}\label{subsec_consistency_criterion}
We now propose a consistency test based on the implied primordial amplitude $ A_s^{\mathrm{Implied}} $. The diagnostic exposes a degeneracy arising while using compressed CMB distance priors to constrain a perturbative degree of freedom decoupled from the background expansion. In such cases, the sampler exploits the $ A_s-D_0 $ degeneracy in \autoref{eq_sigma80} to mimic growth history by inflating $ \sigma_{80} $.

Let $ H(z) $ be the Hubble parameter corresponding to any modified theory of gravity under a spatially flat FLRW metric, then we impose two conditions
\begin{enumerate}
    \item \textit{High redshift $ \Lambda $CDM recovery}:
    \begin{equation}\label{eq_lcdmrecovery}
        H(z) \longrightarrow H_{\Lambda CDM}  \;\;\forall z\geq z_{rec}\simeq1100 \;,
    \end{equation}
    thus preserving the CMB sound horizon, angular diameter distance and damping scale.
    \item \textit{Perturbation modification decoupled from background through $ \lambda $}: \\
    $ \lambda $ enters only in the linear perturbation sector in the growth equation
    \begin{equation}\label{eq_growth}
        \delta_m'' + \delta_m'\left( 2 + \frac{H'}{H} \right) - \frac{3}{2}\Omega_m\mu_G(z, \lambda)\delta_m = 0\;,
    \end{equation}
    where the primes denote derivatives with respect to the e-folding number $ N=\ln a $. The factor $ \mu_G(z, \lambda) = G_{eff}(z, \lambda) / G_N $ obeys 
    \begin{equation}
        \mu_G(z\gg1, \lambda) = 1\;,\label{eq_con2_1}
    \end{equation}
    preserving the high redshift $ \Lambda $CDM recovery from condition 1.
\end{enumerate}
\textit{Corollary}: When satisfying the above conditions, application of the compressed CMB distance priors is valid, since both the background and perturbation physics for $ z\geq z_{rec} $ is governed by $ \Lambda $CDM. The Planck 2018 constraints on the primordial amplitude, $ \ln(10^{10}A_s) = 3.044\pm0.014 $ \cite{Planck:2018vyg}, derived from the same chains as the distance priors, acts as an anchor against which a late-time fit might be tested. This is not a model-independent anchor and its assumptions are discussed in \autoref{subsec_datasets}. To construct the test we begin with the normalization of the linear matter power spectrum \cite{Planck:2013lkt}
\begin{equation}\label{eq_norm_lin_matter_PS}
    \sigma^2(z, R) = \int\frac{k^2 dk}{2\pi^2} P(k, z) |\mathscr{W}(kR)|^2\;.
\end{equation}
The present day clustering amplitude $ \sigma_{80} = \sigma(0, 8h^{-1}Mpc) $, can be further simplified using $ P(k,0)=A_s k^{n_s} T^2(k,\theta) D^2_0(\theta, \lambda) $ and $ \mathcal{I}(\theta)=\int\frac{k^2 dk}{2\pi^2}k^{n_s}T^2(k,\theta) |\mathscr{W}(k\cdot 8h^{-1}Mpc)|^2 $,
\begin{equation}\label{eq_sigma80}
    \sigma_{80}^2 = A_s D^2_0(\theta, \lambda) \mathcal{I}(\theta)\;.
\end{equation}
Here, $ \theta = (\Omega_{m0}h^2, \Omega_bh^2, n_s, h) $ is the parameter vector entering the transfer function and $ D_0 $ is the total growth today\footnote{not to be confused with the normalized ratio $ D(z)/D_0 $ used for $ f\sigma_8 $}. \autoref{eq_sigma80} is invariant under the rescaling $ A_s\rightarrow\varphi^2A_s \;;$ $ D_0\rightarrow D_0/\varphi $ which is the $ A_s-D_0(\lambda) $ degeneracy. This rescaling invariance is exact for the present day, but becomes approximate for redshift dependent likelihoods. The growth data measures $f\sigma_8(z)=\sigma_{80}f(z)\delta_m(z)/\delta_m(0)$ which does not retain its shape under the rescaling, since $\mu_G(z,\lambda)$ modifies both $f(z)$ and the growth factor. The $E_g$ statistic is controlled directly by the background-inert coupling $\lambda$ through $\mu_G(z,\lambda)$ and is independent of the amplitude. Individually, both of these likelihoods break the degeneracy partially, opening a finite $A_s-D_0(\lambda)$ and hence, $\sigma_{80}-\lambda_0$ valley for the sampler to exploit. We quantify this with the posterior correlation of the $\sigma_{80}-\lambda_0$ plane in \autoref{subsec_results2}. Compressed CMB priors constrain distances but leave $ A_s $ and hence $ \varphi $ free, allowing the sampler to absorb this factor into $ \sigma_{80} $ at little statistical cost. When the full CMB likelihood is used, however, the primordial amplitude absorbs this modified growth history, breaking the degeneracy. To realize the impact of $ \varphi $, we use
\begin{equation}\label{eq_sigma80_ref}
    [\sigma_{80}^{\mathrm{ref}}(\theta)]^2 = A_s^{\mathrm{ref}} [D_0^{\mathrm{\Lambda CDM}}]^2 \mathcal{I}(\theta)
\end{equation}
tabulated with a Boltzmann solver over $ \theta $. Taking the ratio of \autoref{eq_sigma80} and \autoref{eq_sigma80_ref} gets us
\begin{multline}\label{eq_asimplied}
    \ln(10^{10}A_s^{\mathrm{Implied}}) = \ln(10^{10}A_s^{\mathrm{ref}}) \\+ 2\ln\left( \sigma_{80}/\sigma_{80}^{\mathrm{ref}} \right) - 2\ln R(\theta, \lambda)
\end{multline}
where the $ \mathcal{I}(\theta) $ cancels out due to the two conditions imposed on the theoretical models, and $ R(\theta, \lambda) = D_0(\theta, \lambda)/D_0^{\Lambda CDM}(\theta) $ is the total growth ratio of the model with the $ \Lambda $CDM reference. The perturbative coupling $ \lambda $ only enters the equation through the growth ratio and hence omitting the latter distorts the very signal to be detected. 

An analogous compensation of the growth rate by the sampler was reported in \cite{Kolhatkar:2026bss} when RSD data was added to the Hybrid $ f(Q) $ pipeline without CMB. The sampler inflated $ \sigma_{80} $ from $ 0.794\pm0.025$ to $ 0.814\pm0.027 $, a $ \sim2.5\% $ shift that was named the amplitude compensation mechanism, but was not traced to its origin. The present work identifies this as a mild instance of the $ A_s-D_0(\lambda) $ degeneracy studied here.
\section{Data and Methodology}\label{sec_DM}
\subsection{Datasets}\label{subsec_datasets}
\begin{enumerate}
    \item \textit{\textbf{Compressed CMB}}: As established by the corollary in \autoref{subsec_consistency_criterion}, all the models considered in this work qualify for applying the compressed CMB data \cite{Zhai:2018vmm}, constraining the vector $ (R, l_a, \Omega_{b0}h^2, n_s) $ where $ h=H_0 / 100 $, and the shift parameters $ R $ and $ l_a $ are defined as
    \begin{gather}
        R = \sqrt{\Omega_{m0} H_0^2}\frac{r(z_*)}{c}\;,\label{eq_shift_R}\\
        l_a = \pi\frac{r(z_*)}{r_s(z_*)}\label{eq_shift_la}\;.
    \end{gather}
    $ r(z_*) = \int_0^{z_*} c\;dz' / H(z')  $ is the comoving distance and $ r_s(z_*) = \int_{z_*}^\infty c_s(z')\;dz' / H(z') $ the comoving sound horizon at the photon decoupling redshift $ z_* $. $ c_s(z) $ is the speed of sound in the photon-baryon fluid. With $ \omega_i = \Omega_{i0}h^2 $, we can write the photon decoupling redshift using the fitting formula \cite{Eisenstein:1997ik} as 
    \begin{equation}\label{eq_zstar}
        z_* = 1048 \left(1+0.00124\omega_b^{-0.738}\right) \left(1+g_1 \omega_m^{g_2}\right)\;,
    \end{equation}
    where 
    \begin{gather}\label{eq_g1g2}
        g_1 = \frac{0.0783\omega_b^{-0.238}}{1+39.5\omega_b^{0.763}}\;,\\
        g_2 = \frac{0.560}{1+21.1\omega_b^{1.81}}\;.
    \end{gather}
    We impose tight Gaussian priors on the parameters $ \omega_b $ and $ n_s $ since they cannot be modified under the present theoretical framework. Both of these parameters are also excluded from the contour plots since their behavior is expected to mimic $ \Lambda $CDM. The covariance matrix for the distance priors is provided in \cite{Zhai:2018vmm}.
    
    \item \textit{$ \bm{E_g} $ \textbf{Statistic}}: We employ the so called $ E_g $ statistic which is a large-scale structure probe introduced in \cite{Zhang:2007nk}, measuring the ratio of the gravitational lensing and matter overdensity. \cite{Ghosh:2018ijm} captures the same statistic in terms of the $ \mu(z,k)-\gamma(z,k) $ parameterization of a modified theory of gravity and the velocity field $ f(z) $ as
    \begin{equation}\label{eq_EG_mugamma}
        E_g(z) = \frac{\Omega_{m0}\mu(z,k)(1+\gamma(z,k))}{2f(z)}\;.
    \end{equation}
    As mentioned in the previous section, the $ f(Q) $ theory in the QSA limit obeys $ \mu(z) = 1/f_Q\;,\;\gamma(z)=1 $ devoid of any scale dependence or anisotropic stress, implying
    \begin{equation}
        E_g(z) = \frac{\Omega_{m0}\mu(z)}{f(z)}\;.
    \end{equation}
    The compilation of \cite{Alestas:2022gcg} contains two VIPERS measurements which are also present in the $ f\sigma_8(z) $ table from the same paper. Both of them are derived from the same PDR-2 catalog and depend on the same growth rate $ f(z) $. Due to the absence of any joint covariance, we remove the two points from the $ E_g $ sample, retaining them in the RSD compilation. This leaves us with 5 uncorrelated points. Retaining the double count shifts $\log\mathcal{Z}$ by up to $\sim9$ nats and was the dominant systematic in an earlier version of this analysis.
    
    \item \textit{\textbf{Type Ia Supernovae}}: The Pantheon$^+$ \cite{Brout:2022vxf, Scolnic:2021amr} compilation consists of 1701 data points measuring the distance modulus
    \begin{equation}
        \mu(z) = 5\log_{10}\left( \frac{d_L(z)}{1Mpc} \right) + 25\;,
    \end{equation}
    out of which we only utilize 1590 points with $ z>0.01 $. We marginalize the absolute magnitude $ \mathcal{M} $. 

    \item \textit{\textbf{Baryon Acoustic Oscillations}}: We use two set of measurements for the BAO observables
    \begin{itemize}
        \item \textbf{SDSS DR16} : We use the BAO+RSD combined measurements released through \cite{eBOSS:2020yzd}. This compilation constrains $ D_V(z)/r_d,\;D_M(z)/r_d,\;D_H(z)/r_d$ and $ f\sigma_8(z) $ spanning $ 0.15\leq z \leq 2.33 $. 
        \item \textbf{DESI DR2 BAO} : We also use the latest DESI DR2 \cite{DESI:2025zgx} release of $ D_V(z)/r_d,\;D_M(z)/r_d$ and $ D_H(z)/r_d $ decoupled from growth data measurements. There are a total of 13 such measurements spanning $ 0.295\leq z\leq2.33 $.
    \end{itemize}
    In the absence of any early time modifications, we keep $ r_d $ fixed. For SDSS, $ r_d = 147.8\;Mpc $ is the fiducial value of the eBOSS analysis and for DESI DR2, $r_d=147.05\;Mpc$ is the output of \texttt{CAMB} for the Planck 2018 best fit with massive neutrinos. A robustness check in \autoref{app_rd} shows that the results do not depend on this choice. An importance reweighting of the chains to either the same $r_d=147.05\;Mpc$ or to a value computed from each $(\omega_b,\omega_m)$, moves $\sigma_{80}$ and $\lambda_0$ by at most $0.1\sigma$ and $0.2\sigma$ respectively.

    \item \textit{\textbf{RSD}}: We use 20 points of isolated $ f\sigma_8(z) $ measurements compiled in \cite{Alestas:2022gcg} as a supplement to the DESI DR2 dataset, since the latter does not contain any growth measurements. 

    \item \textit{\textbf{Planck prior on $ A_s $}}: Finally, we employ a Gaussian prior for $ \ln(10^{10}A_s) $ from Planck 2018 \citep{Planck:2018vyg} for the TT, TE, EE + low E + lensing case as a Gaussian distribution $ \mathcal{N}[3.044,0.014] $. This prior is applied on the derived quantity $ A_s^{\mathrm{implied}} $ from \autoref{eq_asimplied} evaluated at every likelihood call. The constraint acts on the joint vector $ (\sigma_{80}, \lambda_0, \Omega_{m0}, H_0, \omega_b, n_s) $ allowing it to break the $ A_s-D_0(\lambda) $ degeneracy. This prior is only applied to the $ +\lambda_0+\ln(A_s) $\footnote{For notational convenience, we write $ \ln(A_s) $ instead of $ \ln(10^{10}A_s) $} variants to fix the degeneracy issue described in \autoref{subsec_consistency_criterion}. It should be noted that this prior comes with two implicit assumptions. First, the Planck constraint is derived for $\Lambda$CDM with CMB lensing. The conditions \autoref{eq_lcdmrecovery} and \autoref{eq_con2_1} ensure that the primary anisotropies at the last scattering surface are those for $\Lambda$CDM, so that the measured amplitude applies to the models considered here. The lensing channel, however, depends on the late-time growth of the specific model. Reweighting the chains with the TT,TE,EE+lowE constraint without the lensing reconstruction from Planck 2018 $\ln(A_s)=3.045\pm0.016$ \citep{Planck:2018vyg} moves $\sigma_{80}$ and $\lambda_0$ by less than $0.03\sigma$. The lensing smoothing of the primary spectra is present in both and requires a full Boltzmann treatment to probe it. Second, the $\ln(A_s)$ constraint is added on top of the compressed vector as an independent prior, even though both are derived from the same Planck chains. We neglect these correlations, making the prior an approximation appropriate for the diagnostic purposes of this work.
\end{enumerate}

\subsection{Statistical Framework}
Our parameter inference rests on the Bayes theorem. For a given model $ \mathscr{M} $,
\begin{equation}\label{eq_bayes}
    \mathcal{P}(\Theta|\mathcal{D}) = \frac{\mathcal{L}(\mathcal{D}|\Theta)\pi(\Theta)}{\mathcal{Z}(\mathcal{D})}\;.
\end{equation}
Here, $ \Theta $ is the set of parameters of $ \mathscr{M} $ given observational data $ \mathcal{D} $. The quantity on the left hand side of \autoref{eq_bayes} is called the \textit{Posterior} of $ \Theta $ given $ \mathcal{D} $, $ \mathcal{L}(\mathcal{D}|\Theta)\equiv\mathcal{L}(\mathcal{D}) $ is the \textit{likelihood} that a certain instance of $ \Theta $ predicts $ \mathcal{D} $, $ \pi(\Theta) $ is the \textit{prior} on $ \Theta $ and $ \mathcal{Z}(\mathcal{D}) $ is the \textit{Bayesian evidence} for the model. Although irrelevant for parameter estimation, the evidence is useful for comparison between models. It is defined as 
\begin{equation}\label{eq_evidence}
    \mathcal{Z}(\mathcal{D}) = \int \mathcal{L}(\mathcal{D}|\Theta)\pi(\Theta) d\Theta\;.
\end{equation}
The Nested Sampling method intrinsically computes Bayesian evidence, which MCMC algorithms do not. We use the recently released publicly available code \texttt{COSMIX} \cite{kolhatkar_cosmix_2026} available at \url{https://github.com/AmeyaKolhatkar/COSMIX} which includes the Boltzmann solver \texttt{CAMB} \cite{Lewis:2026mif, Lewis:1999bs}, the dynamic nested sampling package \texttt{dynesty} \cite{Speagle:2019ivv} as well as \texttt{GetDist} \cite{Lewis:2019xzd} to visualize posterior chains. The inference is based on the log-likelihood, which is expressed as
\begin{equation}
    \ln\mathcal{L} (\mathcal{D}) \sim -\frac{1}{2}  (\Delta\mathcal{X})^T\mathcal{C}^{-1}\Delta\mathcal{X}\;,
\end{equation}
where $ \Delta\mathcal{X} $ is the difference vector between data and theory, while $ \mathcal{C}^{-1} $ is the inverse of the covariance matrix for that particular dataset. The priors for our analysis are displayed in \autoref{table_priors}
\begin{table}[htbp]
    \centering
    \begin{tabular}{>{\centering\arraybackslash}m{8em}>{\centering\arraybackslash}m{8em}}
    \hline
        \textbf{Parameter} & \textbf{Priors} \\
        \hline
        $ H_0 $ & $ \mathcal{U}[\;50, 90\;] $\\
        $ \Omega_{m0} $ & $ \mathcal{U}[\;0.1, 0.9\;] $\\
        $ \sigma_{80} $ & $ \mathcal{U}[\;0.6,1.2\;] $\\
        $ \lambda_0 $ & $ \mathcal{U}[\;-5.0, 5.0\;] $\\
        $ \alpha_2 $ & $ \mathcal{U}[\;0,5\;] $\\
        \hline
    \end{tabular}
    \caption{$ \mathcal{U}[a,b] $ denotes a flat prior with the upper bound $ b $ and lower bound $ a $.}
    \label{table_priors}
\end{table}
\section{Results}\label{sec_results}
\begin{figure*}
    \centering
    \includegraphics[width=0.49\textwidth]{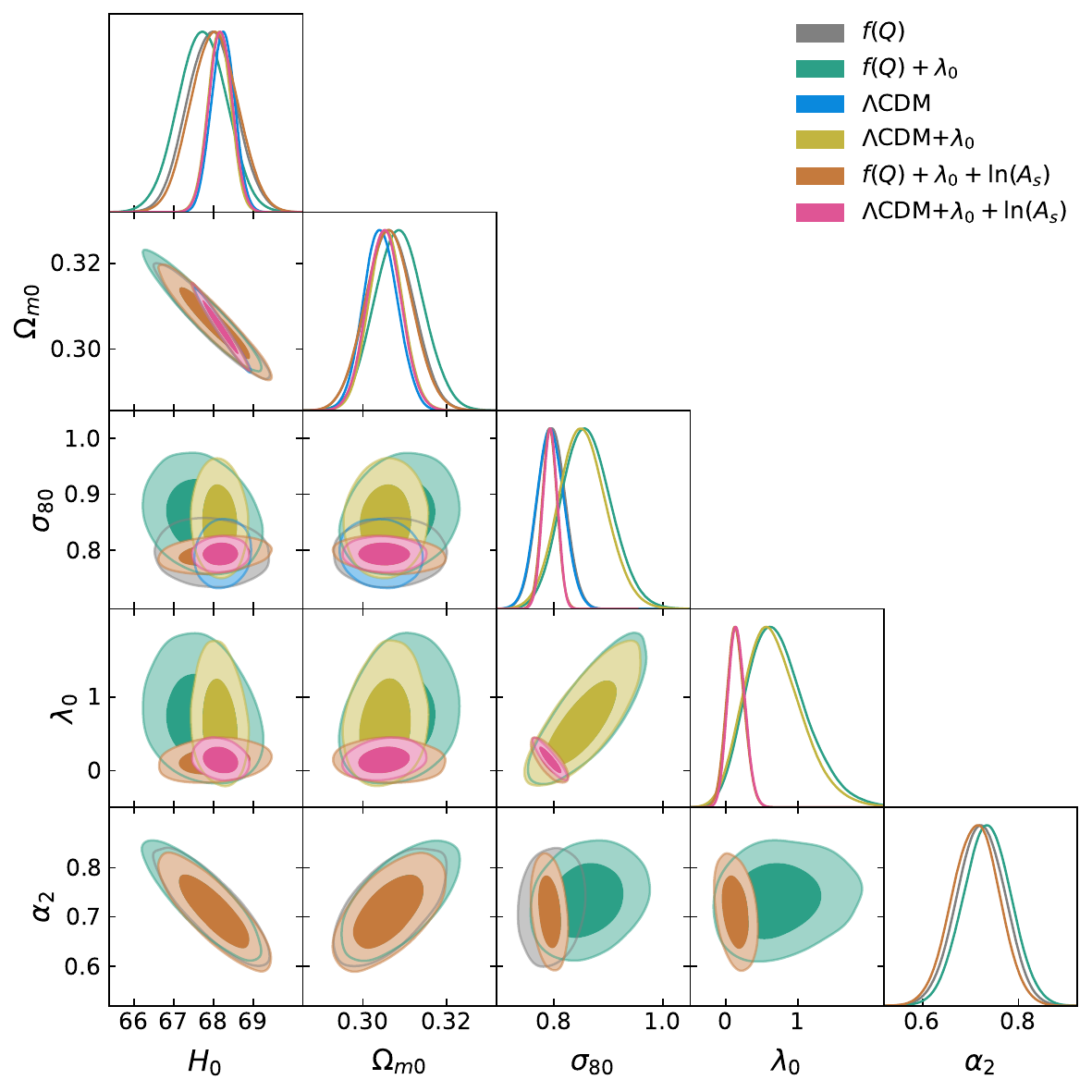}
    \includegraphics[width=0.49\textwidth]{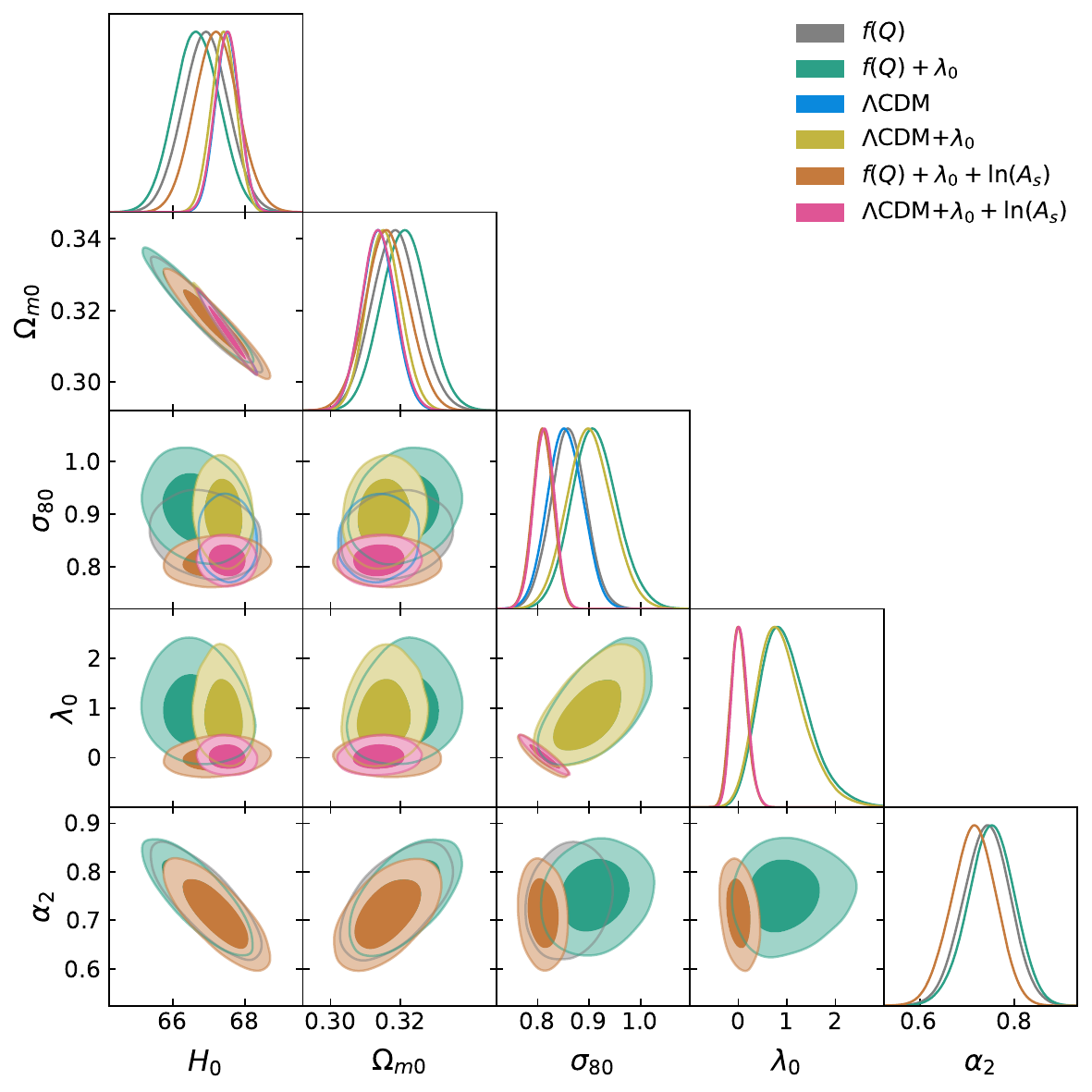}
    \caption{Posteriors for \textit{BD2R} (left panel) and \textit{BSDp} (right panel) combinations}
    \label{fig_corner}
\end{figure*}

This section establishes and demonstrates the $ A_s-D_0(\lambda) $ degeneracy. We build two analysis pipelines out of the \textit{Base} combination - \textit{Compressed CMB + $ E_g $ + Pantheon$ ^+ $ } to maximize the constraining power of the given data. 
\begin{enumerate}
    \item \textit{BD2R : Base + DESI DR2 BAO + RSD}
    \item \textit{BSDp : Base + SDSS DR16 BAO}
\end{enumerate}
The visualized posteriors of chains corresponding to the above combinations are displayed in \autoref{fig_corner} respectively, while \autoref{table_ICE} summarizes the $ 1\sigma $ best-fit values along with Information Criterion and Evidence (ICE) metrics.
\subsection{\textit{Baseline Models}}\label{subsec_results1}
\begin{figure}
    \centering
    \includegraphics[width=0.95\columnwidth]{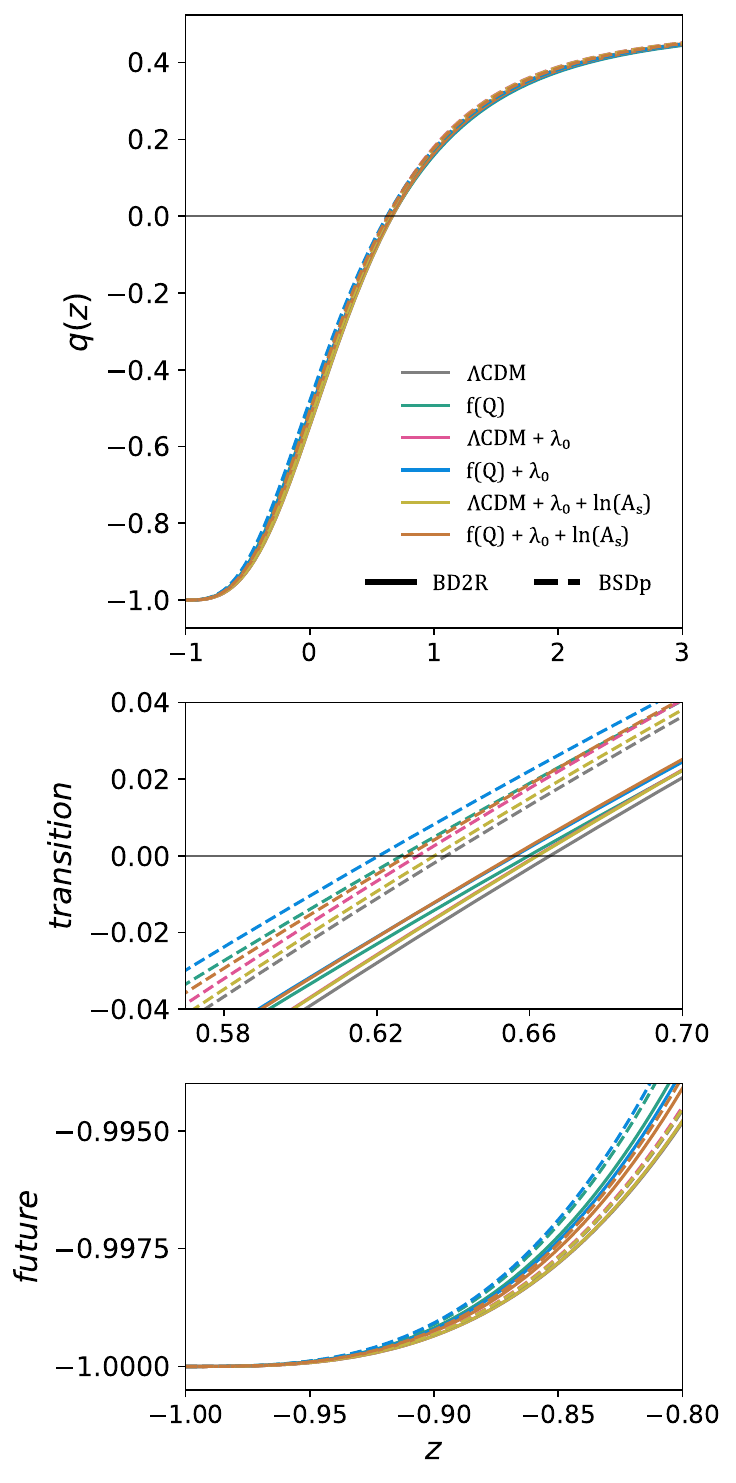}
    \caption{Overlay plot for $ q(z) $ for all the model and dataset combinations}
    \label{fig_q_all}
\end{figure}

\begin{figure}
    \centering
    \includegraphics[width=0.95\columnwidth]{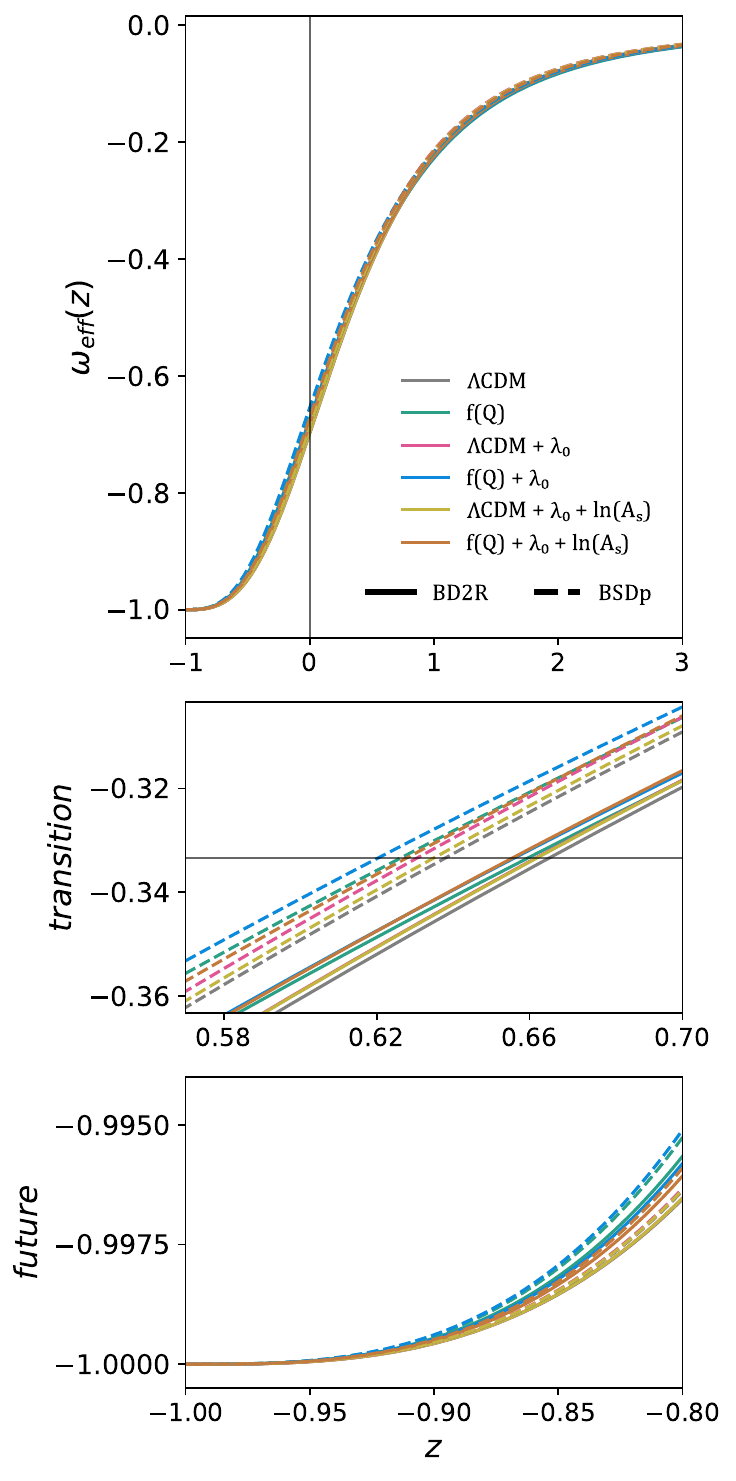}
    \caption{Overlay plot for $ \omega_{eff}(z) $ for all the model and dataset combinations}
    \label{fig_weff_all}
\end{figure}

The parameter vector $ (H_0, \Omega_{m0}, \sigma_{80}) $ remains consistent with standard results for both the models and dataset combinations. The correlation between $ \alpha_2 $ and $ \sigma_{80} $ in the Hybrid model is practically non-existent. This is expected because for the best-fit $ \alpha_2\approx0.7 $, the combination $ 1-\Tilde{\Omega}_0\approx0 $, suppressing the $ f(Q) $ correction in \autoref{eq_muG_fQ} and hence decoupling $ \alpha_2 $ from the growth sector. It is clear from \autoref{table_ICE} that ICE values penalize the Hybrid model for the extra parameter $ \alpha_2 $, which acts as an early time cosmological constant. Using the Jeffrey's scale interpretation \cite{Trotta:2008qt}, we find a weak evidence for the $ \Lambda $CDM model. The background kinematic evolution is summarized in \autoref{table_presentdayvalues} and illustrated in \autoref{fig_q_all} and \autoref{fig_weff_all}, showing the deceleration parameter $ q(z) $ and the effective equation of state parameter $ \omega_{eff}(z) $ respectively. All models recover a matter dominated past ($ q\rightarrow0.5 $ and $ \omega_{eff}\rightarrow0 $) and transition to accelerated expansion at $ z_t\sim0.62-0.67 $ with model-to-model differences below $ 2\% $ within each pipeline.
\begin{table*}
    \centering
    \renewcommand{\arraystretch}{1.5}
    \begin{tabular}{ >{\centering\arraybackslash}m{10em} 
                     >{\centering\arraybackslash}m{6em} 
                     >{\centering\arraybackslash}m{6em} 
                     >{\centering\arraybackslash}m{7em} 
                     >{\centering\arraybackslash}m{6em}
                     >{\centering\arraybackslash}m{6em}
                     >{\centering\arraybackslash}m{7em}}
        \hline
        \textbf{Parameter} & $ \bm{\Lambda} $\textbf{CDM} & $ \bm{\Lambda} $\textbf{CDM} $ \bm{+\lambda_0} $ & $ \bm{\Lambda} $\textbf{CDM} $ \bm{+ \lambda_0 + \ln(A_s)}^\dagger $ & $ \bm{f(Q)} $ & $ \bm{f(Q) + \lambda_0} $ & $ \bm{f(Q) + \lambda_0 + \ln(A_s)}^\dagger $ \\
        \hline
        \multicolumn{7}{c}{\textbf{\textit{BD2R}}}\\
        \hline
        $ \bm{H_0} $ & $ 68.23\pm0.29 $ & $ 68.14\pm0.29 $ & $ 68.17\pm0.29 $ & $ 67.94\pm0.59 $ & $ 67.72\pm0.61 $ & $ 68.01\pm0.58 $ \\
        
        $ \bm{\Omega_{m0}} $ & $ 0.304\pm0.004 $ & $ 0.305\pm0.004 $ & $ 0.305\pm0.004 $ & $ 0.306\pm0.006 $ & $ 0.309\pm0.006 $ & $ 0.306\pm0.006 $ \\
        
        $ \bm{\sigma_{80}} $ & $ 0.794\pm0.025 $ & $ 0.853^{+0.040}_{-0.045} $ & $ 0.793\pm0.013 $ & $ 0.795\pm0.025 $ & $ 0.861^{+0.041}_{-0.048} $ & $ 0.792\pm0.014 $ \\
        
        $ \bm{\lambda_0} $ & --------- & $ 0.67^{+0.32}_{-0.46} $ & $ 0.15^{+0.11}_{-0.13} $ & --------- & $ 0.73^{+0.33}_{-0.49} $ & $ 0.13\pm0.12 $ \\
        
        $ \bm{\alpha_2} $ & --------- & --------- & --------- & $ 0.722\pm0.050 $ & $ 0.734\pm0.050 $ & $ 0.711\pm0.049 $ \\
        \hline
        $ \bm{\Delta} $\textbf{AIC} & $ 0 $  & $ -0.5 $ & $ +1.1 $ & $ +1.2 $ & $ +0.2 $ & $ +2.7 $ \\
        
        $ \bm{\Delta} $\textbf{BIC} & $ 0 $ & $ +4.9 $ & $ +6.5 $  & $ +6.6 $ & $ +11.0 $ & $ +13.5 $ \\
        
        $ \bm{\Delta} $\textbf{DIC} & $ 0 $ & $ -0.7 $ & $ +1.1 $  & $ +1.6 $ & $ +0.5 $ & $ +2.7 $ \\
        
        $ \bm{\Delta\log\mathcal{Z}} $ & $ 0 $ & $ -1.40\pm0.14 $ & --------- & $ -1.63\pm0.15 $ & $ -2.60\pm0.15 $ & --------- \\
        $\bm{\ln\mathcal{P}(A_s^{\rm Planck}|\mathcal{D},\mathcal{M})}$ & --------- & --------- & $-3.44\pm0.16$ & --------- & --------- & $-3.75\pm0.17$ \\
        \hline
        \multicolumn{7}{c}{\textbf{\textit{BSDp}}}\\
        \hline
        $ \bm{H_0} $ & $ 67.51\pm0.33 $ & $ 67.40\pm0.34 $ & $ 67.50\pm0.34 $ & $ 66.91\pm0.63 $ & $ 66.67\pm0.63 $ & $ 67.20\pm0.60 $ \\
        
        $ \bm{\Omega_{m0}} $ & $ 0.314\pm0.005 $ & $ 0.315\pm0.005 $ & $ 0.314\pm0.005 $ & $ 0.319\pm0.007 $ & $ 0.321\pm0.007 $ & $ 0.316\pm0.006 $ \\
        
        $ \bm{\sigma_{80}} $ & $ 0.854\pm0.034 $ & $ 0.901\pm0.044 $ & $ 0.812\pm0.020 $ & $ 0.860\pm0.034 $ & $ 0.911^{+0.041}_{-0.047} $ & $ 0.810\pm0.020 $ \\
        
        $ \bm{\lambda_0} $ & --------- & $ 0.90^{+0.37}_{-0.58} $ & $ 0.03^{+0.15}_{-0.18} $ & --------- & $ 0.97^{+0.39}_{-0.59} $ & $ 0.01^{+0.15}_{-0.18} $ \\
        
        $ \bm{\alpha_2} $ & --------- & --------- & --------- & $ 0.741\pm0.049 $ & $ 0.751\pm0.049 $ & $ 0.714\pm0.047 $ \\
        \hline
        $ \bm{\Delta} $\textbf{AIC} & $ 0 $  & $ -1.7 $ & $ +3.4 $ & $ +0.6 $ & $ -1.5 $ & $ +5.0 $ \\
        
        $ \bm{\Delta} $\textbf{BIC} & $ 0 $ & $ +3.7 $ & $ +8.8 $ & $ +6.0 $ & $ +9.2 $ & $ +15.8 $ \\
        
        $ \bm{\Delta} $\textbf{DIC} & $ 0 $ & $ -1.7 $ & $ +3.3 $ & $ +0.7 $ & $ -1.9 $ & $ +4.7 $ \\
        
        $ \bm{\Delta\log\mathcal{Z}} $ & $ 0 $ & $ -0.48\pm0.14 $ & --------- & $ -1.86\pm0.14 $ & $ -1.46\pm0.14 $ & --------- \\
        $\bm{\ln\mathcal{P}(A_s^{\rm Planck}|\mathcal{D},\mathcal{M})}$ & --------- & --------- & $-5.68\pm0.15$ & --------- & --------- & $-6.09\pm0.16$ \\
        \hline
    \end{tabular}
    \caption{ICE Table for \textit{BD2R} and \textit{BSDp}. All $\Delta$ values are referenced to vanilla $\Lambda$CDM within the same pipeline. $^\dagger$These variants include the Planck amplitude prior as an additional likelihood term, so their evidence is computed under a different dataset than the reference. The corresponding $\Delta\log\mathcal{Z}$ is therefore not a Bayes factor and thus, the expression \autoref{eq_predictive} in \autoref{subsec_firewall} is used and reported as a separate metric. The uncertainties on the evidence values are \texttt{dynesty} estimates and are carried over to predictive density values as well. The information criteria for these variants are taken with respect to $\Lambda$CDM without the prior, which only contributes at most $0.13$, to $\chi^2_{min}$ at the best-fit.
    }
    \label{table_ICE}
\end{table*}
\subsection{\textit{The \texorpdfstring{$ \lambda_0 $}{Lg} coupling}}\label{subsec_results2}
\begin{figure}
    \centering
    \includegraphics[width=0.85\columnwidth]{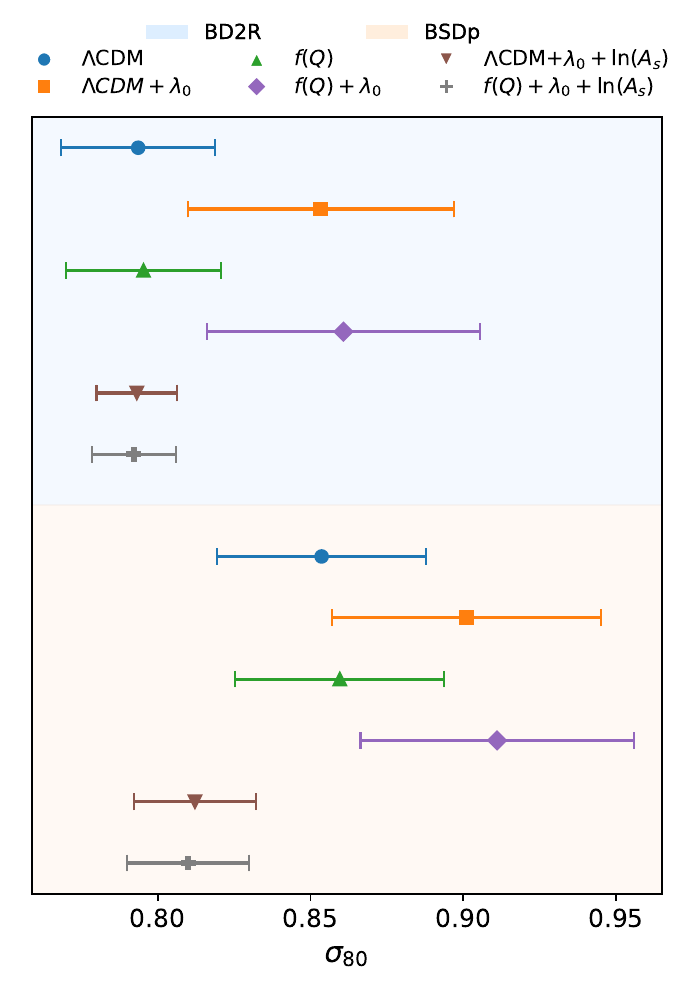}
    \caption{Whisker plot for $ \sigma_{80} $ for all the model and dataset combinations}
    \label{fig_whisker}
\end{figure}
\begin{figure}
    \centering
    \includegraphics[width=0.95\columnwidth]{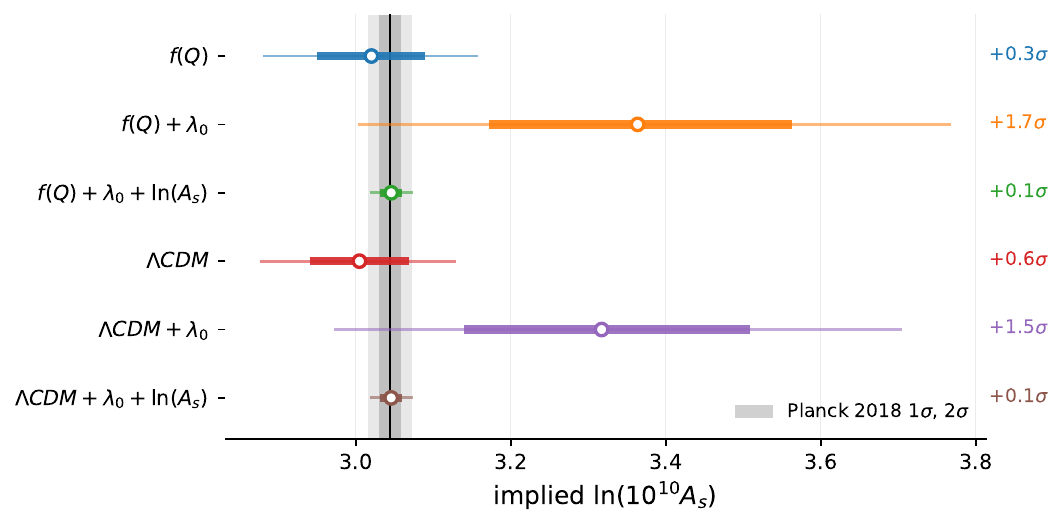}
    \caption{The implied $ A_s $ for \textit{BD2R}}
    \label{fig_impliedAs_BD2R}
\end{figure}
\begin{figure}
    \centering
    \includegraphics[width=0.95\columnwidth]{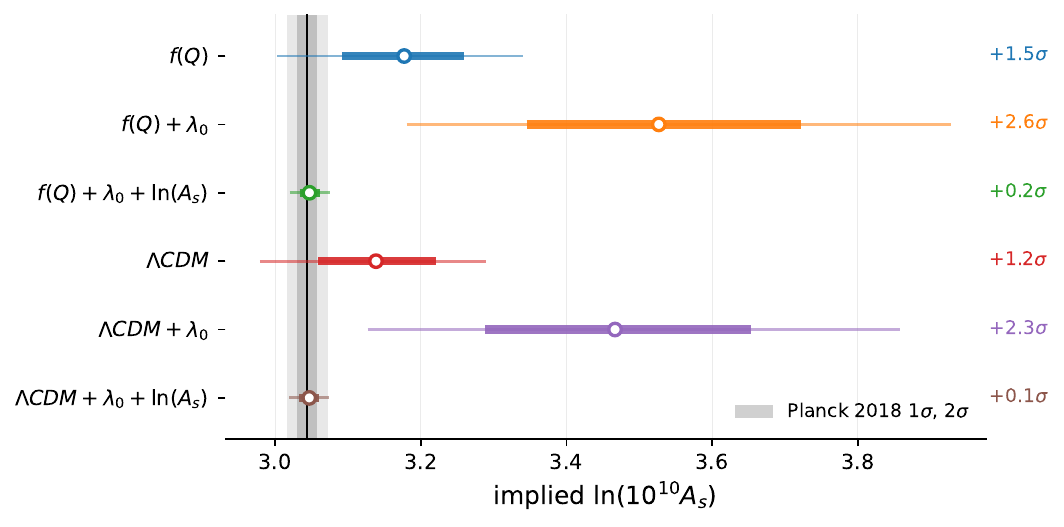}
    \caption{The implied $ A_s $ for \textit{BSDp}}
    \label{fig_impliedAs_BSDp}
\end{figure}
\begin{figure*}
    \centering
    \includegraphics[width=0.95\textwidth]{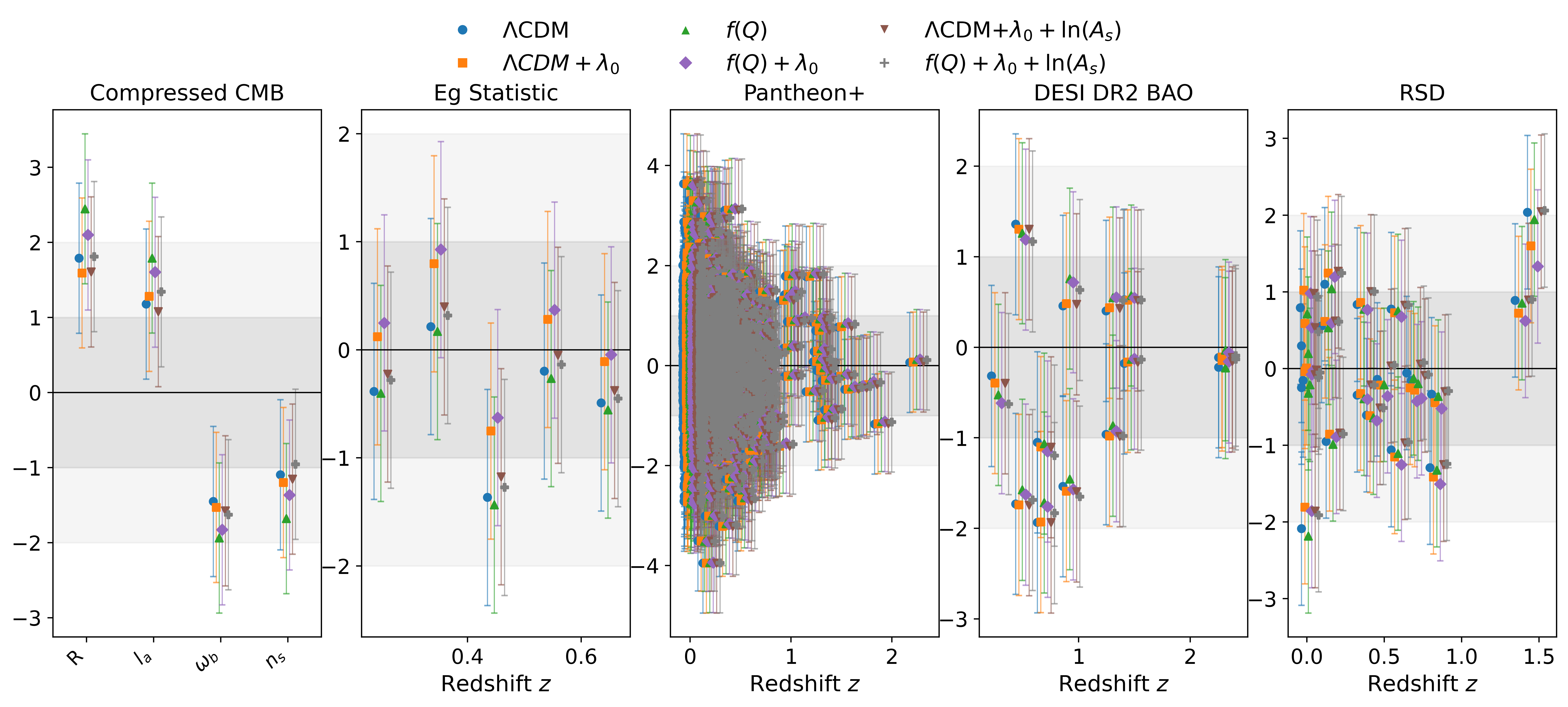}
    \caption{Pull plot for \textit{BD2R}}
    \label{fig_pull_plot_BD2R}
\end{figure*}
\begin{figure*}
    \centering
    \includegraphics[width=0.95\textwidth]{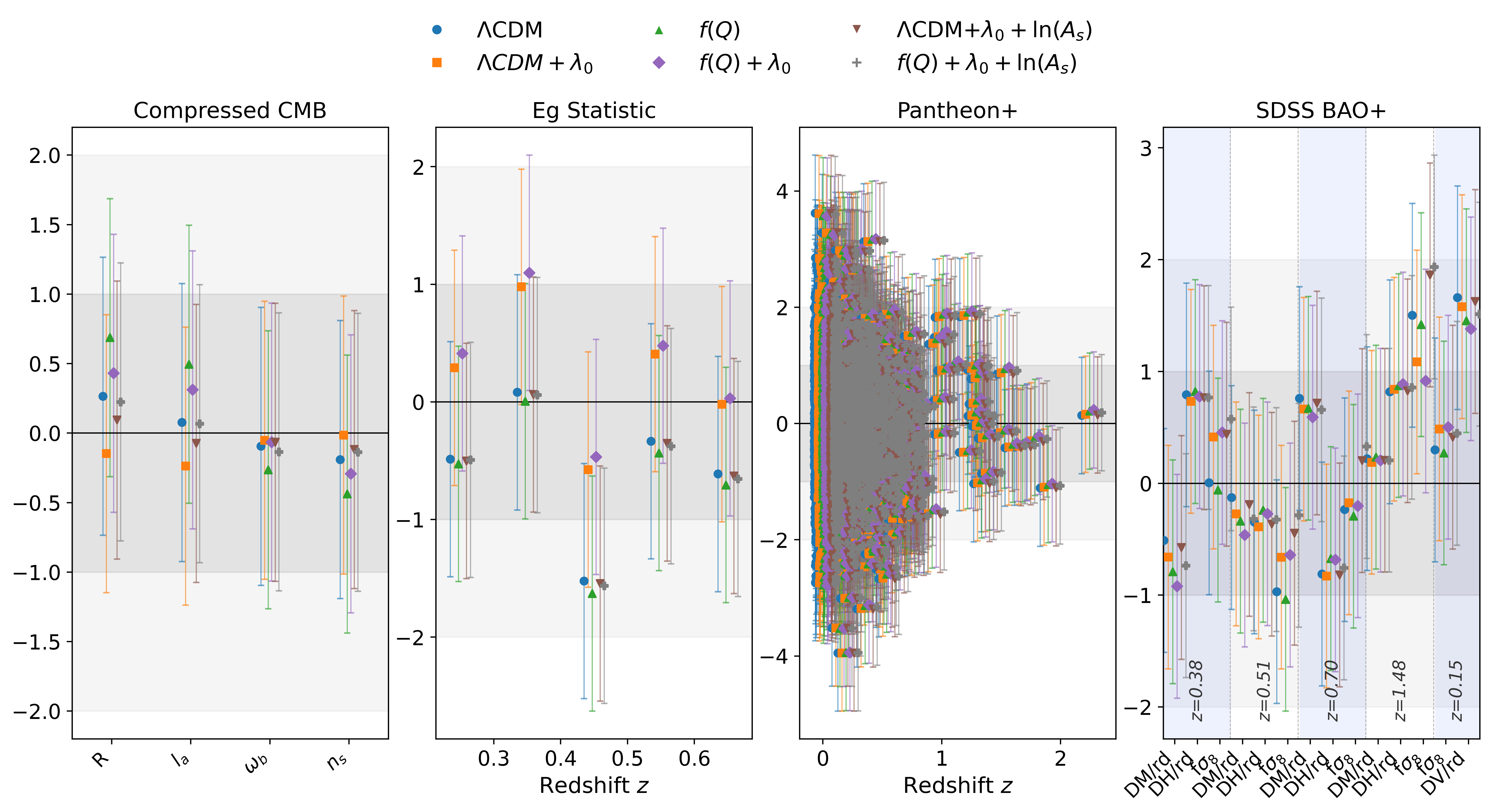}
    \caption{Pull plot for \textit{BSDp}}
    \label{fig_pull_plot_BSDp}
\end{figure*}

When the background-inert parameter $ \lambda_0 $ is added, the $ \sigma_{80} $ contours widen, and show a positive correlation with $ \lambda_0 $, with the correlation coefficient of $ +0.78 $ and $ +0.79 $ on \textit{BD2R} and $ +0.57 $ and $ +0.59 $ on \textit{BSDp} for the $ \Lambda $CDM and Hybrid backgrounds respectively. This is an illustration of the $ A_s-D_0(\lambda) $ degeneracy discussed in \autoref{subsec_consistency_criterion} manifest as $ \sigma_{80}-\lambda_0 $ degeneracy. When allowed to vary unrestricted, the sampler will choose a suitable growth factor, and hence $ \sigma_{80} $, to accommodate the given data and ``slide" along the $ \sigma_{80}-\lambda_0 $ valley in search of a statistically preferred point. This drives the ICE values to be in mutual disagreement for the $ +\lambda_0 $ variants. \autoref{fig_whisker} shows the $ \sigma_{80} $ whisker plot for all the model and dataset combinations, directly visualizing the systematic upward shift induced by the $ \lambda_0 $ coupling, and its resolution upon imposing the $ \ln(A_s) $ prior. Further, \autoref{fig_impliedAs_BD2R} and \autoref{fig_impliedAs_BSDp} demonstrate this through the implied $ A_s $ given by \autoref{eq_asimplied}. The values reported by these figures are computed using the standard formula
\begin{equation}\label{eq_tension}
    t = \frac{|m_1-m_2|}{\sqrt{\sigma_1^2+\sigma_2^2}}\;,
\end{equation}
with the two quantities represented by Gaussian distributions $\mathcal{N}[m_i,\sigma_i]$, such that $i=1$ corresponds to the mean and standard deviation of the implied $\ln(A_s)$ and $i=2$ corresponds to $\mathcal{N}[3.044,0.014]$. The $ \lambda_0 $ coupling drives $ A_s^{\mathrm{implied}} $ to $ 1.32-1.38 $ times the Planck values for \textit{BD2R} and $ 1.53-1.63 $ for \textit{BSDp}. This cements the resourcefulness of the implied $ A_s $ consistency test, wherein a simple check can prevent misinference motivated by inconclusive evidence or inconsistency between various evidence metrics. The pull plots for the \textit{BD2R} and \textit{BSDp} combinations are displayed in \autoref{fig_pull_plot_BD2R} and \autoref{fig_pull_plot_BSDp} respectively. The residuals across all individual datasets -- Compressed CMB, $ E_g $ statistic, Pantheon$ ^+ $, BAO and RSD -- are statistically comparable between the $ +\lambda_0 $ variants and their baselines, confirming that the $ \chi^2_{min} $ improvement is not driven by an improved fit to any single observable but arises globally from the degeneracy inflated amplitude.
\subsection{Planck prior on \texorpdfstring{$ \ln(A_s) $}{Lg}}\label{subsec_results3}
Imposing the $ \ln(A_s) $ priors from Planck causes the contours to constrict into even smaller shapes than in \autoref{subsec_results1}, breaking the $ A_s-D_0(\lambda) $ degeneracy. When the degeneracy breaks, the $ A_s^{\mathrm{implied}} $ values fall within $ 0.2\sigma $ of Planck across combinations and the $ \lambda_0 $ coupling becomes consistent with $ 0 $. Additionally, the $ \sigma_{80}-\lambda_0 $ contours become negatively correlated due to the fixed primordial amplitude. We do not quote evidence ratios for these variants. The amplitude prior enters as an additional likelihood term, so their evidence is computed under a different dataset (see \autoref{table_ICE} and \autoref{subsec_firewall}). What the comparison does support is stated there in terms of the predictive density of the Planck amplitude measurement.
\subsection{Evaluation of \texorpdfstring{$\varepsilon_{sc}$}{Lg}}\label{subsec_results4}
\begin{figure}
    \centering
    \includegraphics[width=0.95\columnwidth]{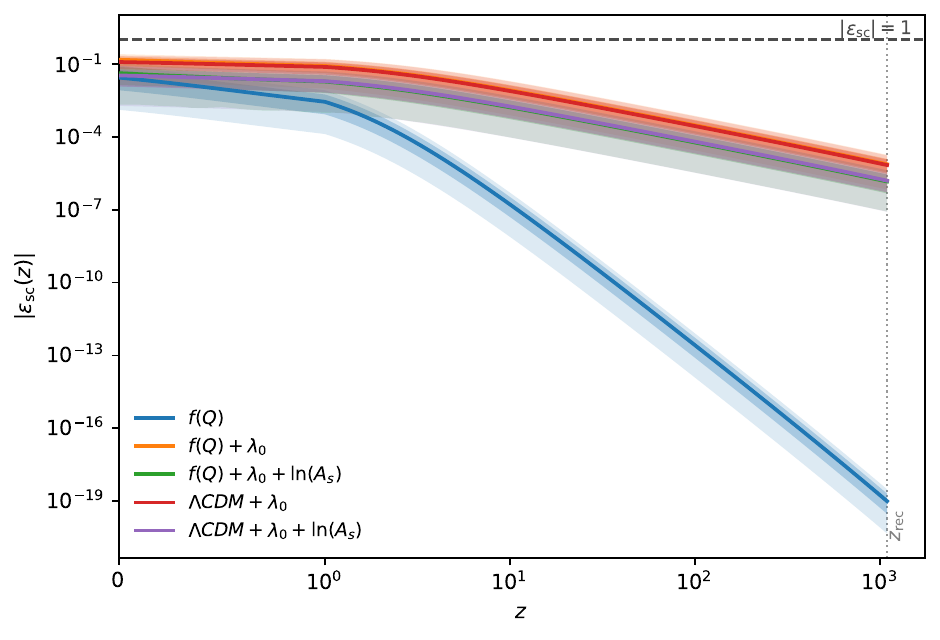}
    \caption{The strong coupling proximity parameter for BD2R combination}
    \label{fig_epsilon_sc_BD2R}
\end{figure}

\begin{figure}
    \centering
    \includegraphics[width=0.95\columnwidth]{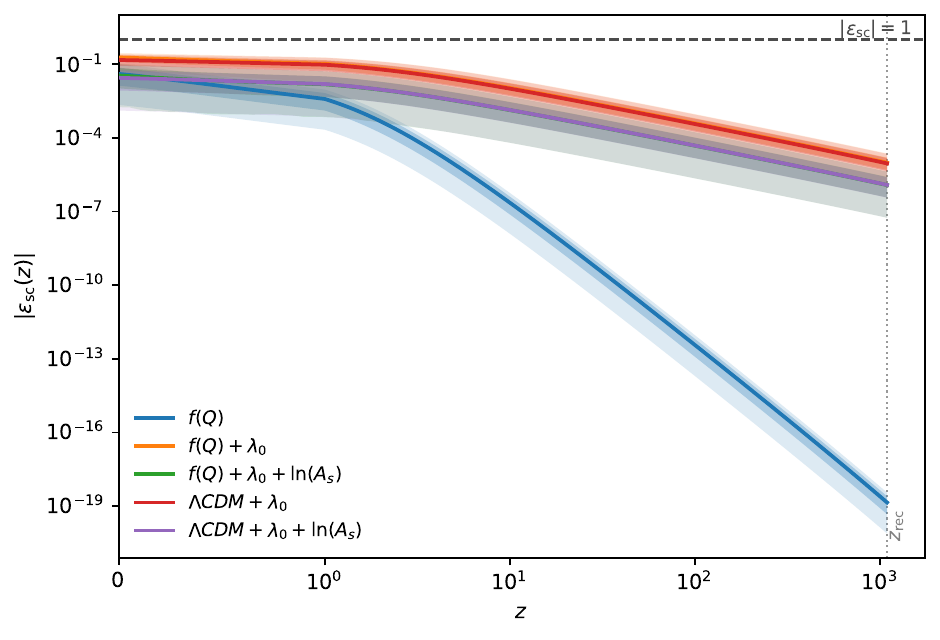}
    \caption{The strong coupling proximity parameter for BSDp combination}
    \label{fig_epsilon_sc_BSDp}
\end{figure}

\autoref{fig_epsilon_sc_BD2R} and \autoref{fig_epsilon_sc_BSDp} show
$ \varepsilon_{sc}(z) $ from recombination to the present. For baseline
$ \Lambda $CDM $ f_{QQ}=0 $ and $ \varepsilon_{sc} $ vanishes
identically, as it must. The $ +\lambda_0 $ variants reach
$ \varepsilon_{sc}(0)=0.120 $ and $ 0.149 $ for \textit{BD2R} and $ 0.149 $ and
$ 0.188 $ for \textit{BSDp} ($ \Lambda $CDM and Hybrid respectively), with $ 95\% $ upper limits of $ 0.22-0.29 $. Imposing the amplitude prior reduces these to
$ 0.028-0.042 $, a factor of $ 3.5-5 $. All values remain well below unity, so no
posterior region approaches the $ f_Q\rightarrow0 $ pathology. We stress
that $ \varepsilon_{sc} $ measures proximity to that boundary and is not a
substitute for a full perturbative stability analysis.
\section{Discussion}\label{sec_discussion}
\subsection{The \texorpdfstring{$ \sigma_{80}-\lambda_0 $}{Lg} degeneracy valley}\label{subsec_degeneracyvalley}
\begin{figure}
    \centering
    \includegraphics[width=0.95\columnwidth]{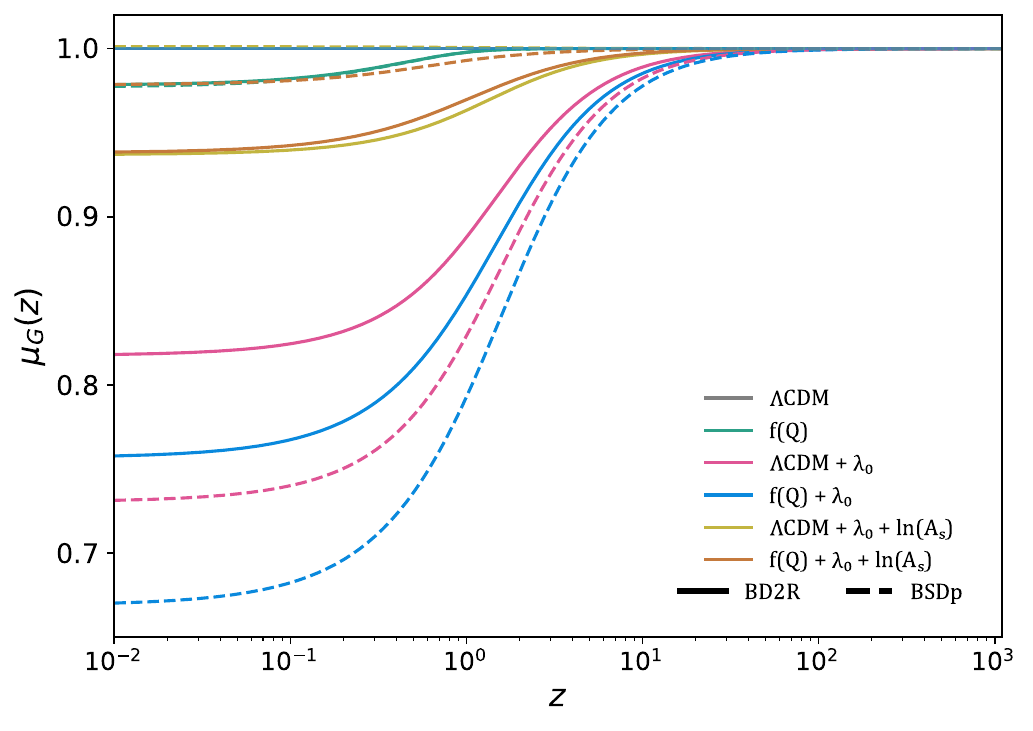}
    \caption{The reconstructed $ \mu_G(z) $ plot for all the model and dataset combinations}
    \label{fig_muG}
\end{figure}

\begin{figure}
    \centering
    \includegraphics[width=0.95\columnwidth]{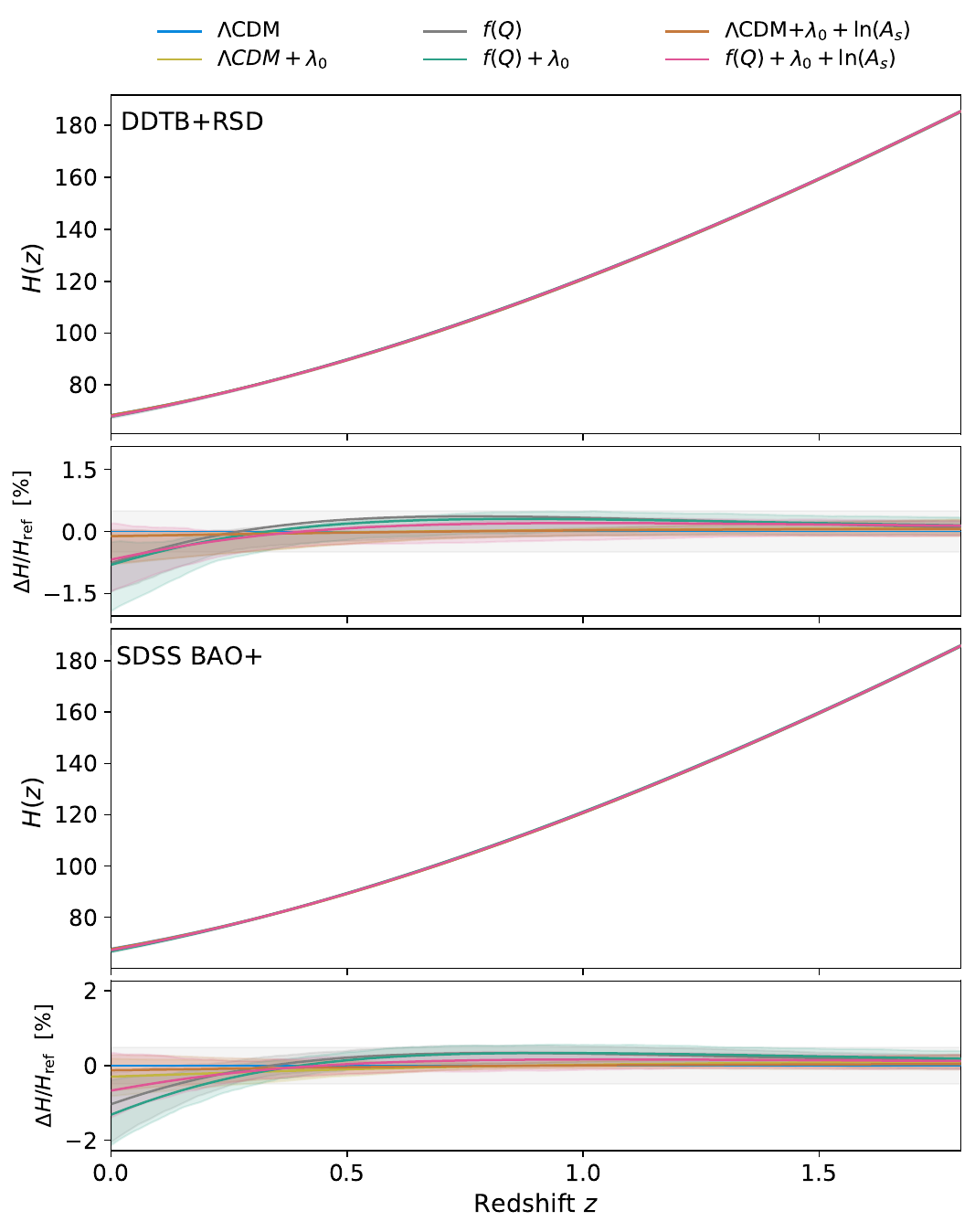}
    \caption{Overlay plot for $ H(z) $ for all the model and dataset combinations}
    \label{fig_H_all}
\end{figure}

\begin{figure}
    \centering
    \includegraphics[width=0.95\columnwidth]{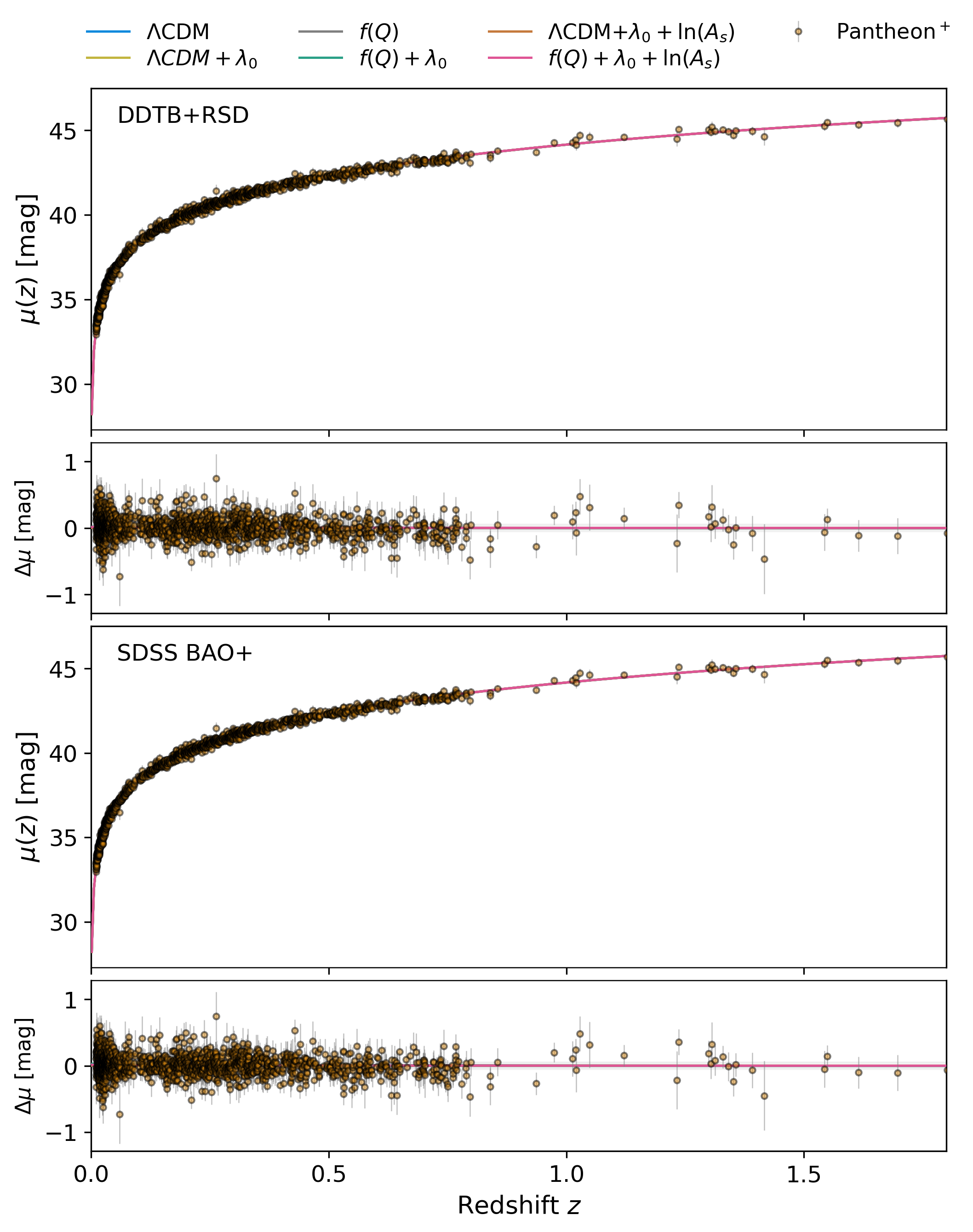}
    \caption{Overlay plot for the distance modulus $ \mu(z) $ for all the model and dataset combinations}
    \label{fig_mu_all}
\end{figure}

\begin{figure}
    \centering
    \includegraphics[width=0.95\columnwidth]{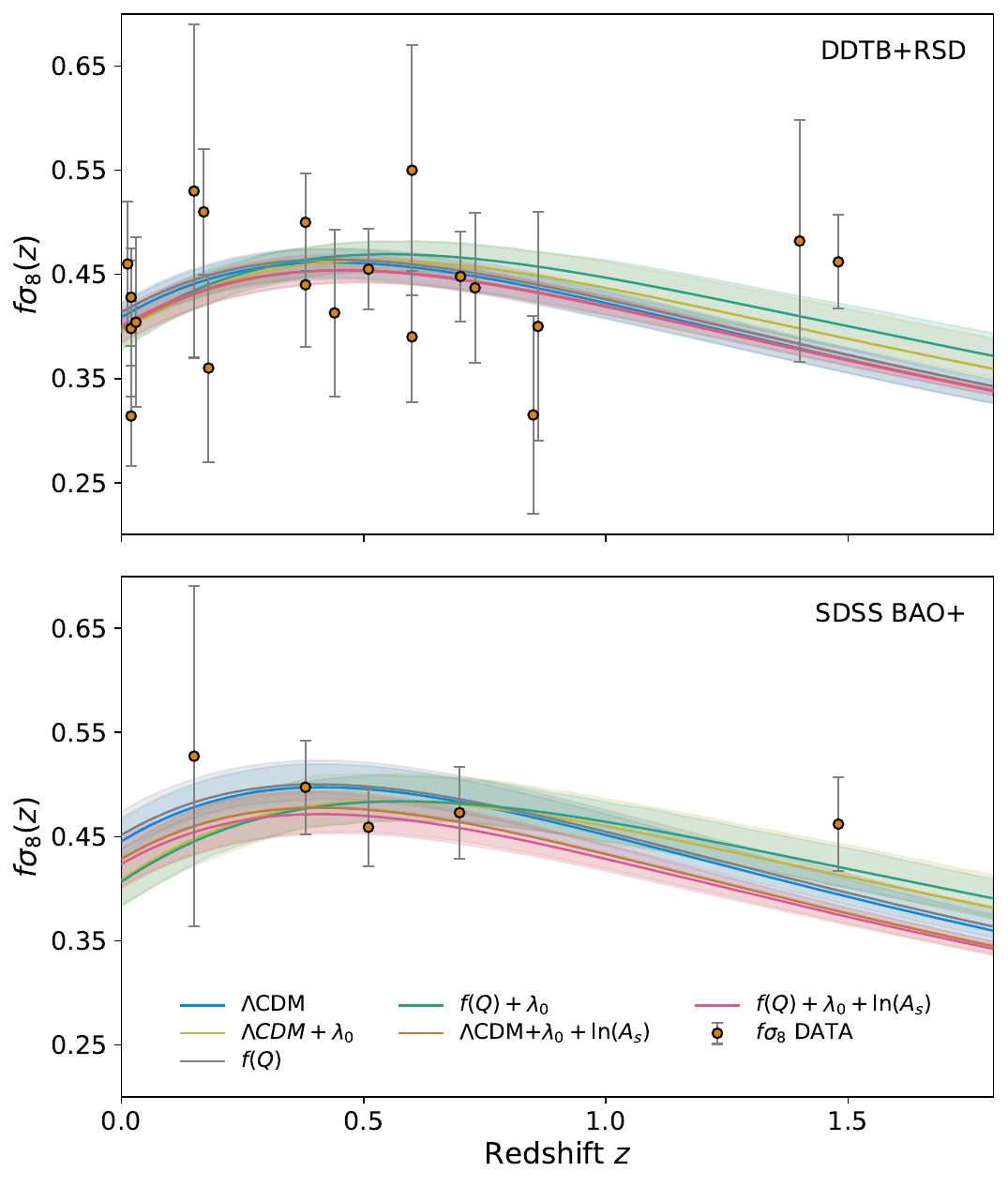}
    \caption{Overlay plot for $ f\sigma_8(z) $ for all the model and dataset combinations}
    \label{fig_fs8_all}
\end{figure}

The reconstructed $ \mu_G(z) $ profile in \autoref{fig_muG} accentuates the physical origin of the observed parameter inflation. $ \mu_G(z) < 1 $ for all the $ \lambda_0>0 $ variants demonstrates that gravity weakens under the $ \sqrt{Q} $ correction. The present day values show maximum suppression of $ 18\%-33\% $ that reduces with increasing redshift, with $ \mu_G\longrightarrow1 $ far before recombination. Crucially, this modification is invisible to the background expansion history, confirmed by \autoref{fig_H_all} and \autoref{fig_mu_all} where all models are consistent with the data in percent level accuracy. This confirms the conditions in \autoref{subsec_consistency_criterion}.
Weaker gravity suppresses the source term in \autoref{eq_growth}, causing the growth rate $ f(z) $ to decrease. Since the observed quantity is actually the product $ f\sigma_8(z) $, the sampler compensates by artificially inflating the $ \sigma_{80} $ value. A higher $ \lambda_0 $ requires a higher compensation, and hence a higher $ \sigma_{80} $. \autoref{fig_fs8_all} makes this compensation explicit. Despite a difference of $ \sim33\% $ in $ \mu_G(0) $ and $ \sim8\% $ in $ \sigma_{80} $ between model variants, the $ f\sigma_8(z) $ curves stay within the error bars of the data. The compensation forms the $ \sigma_{80}-\lambda_0 $ degeneracy valley along which the sampler finds multiple pairs of $ (\sigma_{80}, \lambda_0) $ with a significant fit to data. Adding the full CMB data will break this degeneracy by probing the clustering amplitude, but the compressed CMB only constrains the acoustic geometry through the shift parameters. Nonetheless, adding the amplitude prior not only breaks this degeneracy, but also flips the nature of correlation from positive to negative with tight contour bounds, exhibiting its effectiveness. 
The amplitude compensation mechanism reported in \cite{Kolhatkar:2026bss} is an observational instance of this degeneracy. As mentioned in \autoref{subsec_consistency_criterion}, the inflation was a mere $ \sim2.5\% $ because: (i) the coupling was the fixed 1/Q term, not a free parameter; and (ii) without any CMB input, the RSD data itself constrained $ \sigma_{80} $ from above. In the current pipeline, both conditions are inverted -- $ \lambda_0 $ is a free parameter with no upper bound, and the compressed CMB only anchors to background geometry, leaving the primordial amplitude unconstrained. This extends the ridge for the sampler to explore further. The quantitative difference ($\sim2.5\%$ there and $ \sim5-8\% $ here) is therefore not a discrepancy but a confirmation that the two pipelines probe the same algebraic structure under distinct conditions. Although the companion study identified the phenomenon, this work identifies the cause and quantifies the boundary conditions under which it becomes observationally dangerous.
The implied-$ A_s $ consistency check provides a falsifiable test to ensure consistency with the early Universe untouched by the models currently in use. For \textit{BD2R}, a $ 7\%-8\% $ increase in $ \sigma_{80} $ corresponds to a $ 32\%-38\% $ increase in $ A_s $ at $ 1.5\sigma-1.7\sigma $ tension with Planck. Interestingly, for \textit{BSDp}, a $ 5\%-6\% $ increase corresponds to a $ 53\%-63\% $ increase in $ A_s $ in $ 2.3\sigma-2.6\sigma $ tension with Planck. This reasserts the importance of including the growth ratio $ R(\theta,\lambda) $ from \autoref{eq_asimplied}, without which, a pure $ \sigma_{80}^{\mathrm{ref}} $ driven diagnostic would have it backwards. In both cases, the verdict on statistical preference does not reach a decisive level even for physically inadmissible points. Finally, the $ S_8 $ column from \autoref{table_presentdayvalues} corresponding to the $ +\lambda_0 $ variants ranges from $ 0.861-0.943 $, which are much higher as compared to Planck. This indicates that a background-inert suppression of gravity cannot relieve the $ S_8 $ tension under Compressed CMB priors, because the amplitude compensates for it.
\subsection{Statistical Evidence Metrics under degeneracy}\label{subsec_statisticalmetricsinvalley}
ICE values in \autoref{table_ICE} reveal a systematic tension between evidence metrics under degeneracy. The $ +\lambda_0 $ variants are mildly favored by AIC and DIC, in three out of four cases. The BIC imposes a heavier $ \ln N $ penalty, pushing the same variants into the weakly to strongly disfavored zone. $ \log\mathcal{Z} $ disfavors across model-dataset combinations with the exception of the $ \Lambda $CDM$ +\lambda_0 $ variant under \textit{BSDp}, where it lies in the inconclusive range. It should be noted that $ \log\mathcal{Z} $ scales with the assumed prior range, which for $ \lambda_0 $ is $ \mathcal{U}[-5,5] $ as per \autoref{table_priors}, which means that a choice of a stricter or a more relaxed prior will be reflected in the Bayes factor values. This calls for a combined reading of all the statistical comparison metrics rather than relying on a single criterion.

The degeneracy genuinely improves the $ \chi^2_{min} $ values against one extra parameter. This is manifest in AIC and DIC, but the improvement is not good enough to overcome the harsher BIC. Bayesian evidence also disfavors the variants in the weak to moderate range. None of these metrics actually respond to the amplitude. The implied $ A_s $ diagnostic does exactly that.
\subsection{The Planck \texorpdfstring{$ \ln(A_s) $}{Lg} Firewall}\label{subsec_firewall}
Imposing Planck $ \ln(A_s) $ priors on $ +\lambda_0 $ variants breaks the degeneracy by anchoring $ \sigma_{80} $ to early Universe physics rather than allowing it to float with growth data. The perturbative coupling $ \lambda_0 $ collapses to being consistent with $ 0 $ in every case. Referring to \autoref{table_ICE}, we see that the medians fall by $ 77\%-82\% $ under \textit{BD2R} and by more than $ 96\% $ for \textit{BSDp}, while the $ 1\sigma $ confidence intervals shrink by $ 66\%-70\% $. All the four posteriors become consistent with $ \lambda_0 = 0 $. The present day gravitational coupling $ \mu_G(0) $ recovers similarly, rising from $ 0.67-0.82 $ to $ 0.93-1.0 $ collectively.

A more direct signature of the amplitude prior action is the sign of the $ \sigma_{80}-\lambda_0 $ correlation. Before the prior, the two parameters are positively correlated, due to the valley described in \autoref{subsec_degeneracyvalley}. Once the prior is imposed, the primordial amplitude is fixed, forcing them to trade against each other instead of sliding along the valley together. This inverts the correlation, where the coefficients go from between $+0.57$ and $0.79$, to between $-0.74$ and $-0.95$. The implied amplitude diagnostic works as intended, and all the four runs return $ \ln(10^{10}A_s) $ within $ 0.2\sigma $ of Planck.

We deliberately do not quote the $ \log\mathcal{Z} $ between $ +\lambda_0 $ and $ \lambda_0+\ln(A_s) $ variants due to an additional prior likelihood in the latter, making them unfit for direct model comparison. However, since the two runs only differ in the presence of the $ \ln(A_s) $ prior,  
\begin{equation}\label{eq_predictive}
    \log\mathcal{Z}_{+\lambda_0+\ln A_s} - \log\mathcal{Z}_{+\lambda_0}
      = \log\mathcal{P}\left( A_s^{\mathrm{Planck}}\middle|\mathcal{D,M} \right)\;.
\end{equation}
This is the log predictive density of the Planck amplitude measurement under each model-dataset combination already fitted to late-time data. We obtain, for \textit{BD2R} $ -3.44 $ \& $ -3.75 $ and for \textit{BSDp}, $ -5.68 $ \& $ -6.09 $ up to the common additive constant $\ln(\sigma_{\ln A_s}\sqrt{2\pi})$, since likelihood normalizations are omitted throughout and only differences are used. From \autoref{subsec_results2} and \autoref{subsec_results3}, the \textit{BSDp} results demanding a $ 53\%-63\% $ increase in $ A_s $ are penalized by about $ 2 $ nats more than the \textit{BD2R} combination, which only demands a $ 32\%-38\% $ inflation in $ A_s $. These two independent routes quantify the same inconsistency arising due to the degeneracy.
\subsection{Scope and Applicability of the diagnostic test}
The conditions stated in \autoref{subsec_consistency_criterion} are sufficient (not necessary) for the application of the implied $ A_s $ diagnostic test. Any theory satisfying conditions \autoref{eq_lcdmrecovery} and \autoref{eq_con2_1} will generate the $ A_s-D_0(\lambda) $ degeneracy under compressed CMB priors. This includes any phenomenological modified cosmology with a coupling solely altering the effective gravitational coupling $ \mu_G(z, \lambda) = G_{eff}(z, \lambda) /G_N$, and completely invisible to the background expansion. Both  \cite{Li:2025msm} and \cite{Kolhatkar:2026bss} investigate the $ \sqrt{Q} $ perturbation correction using RSD data and find a mild amplification of $ \sigma_{80} $ in each case; \cite{Li:2025msm} explicitly notes a residual degeneracy between $ M $ and $ \sigma_8 $ but does not trace its origin, while \cite{Kolhatkar:2026bss} names it the amplitude compensation mechanism without diagnosing the cause. Neither study employs compressed CMB and hence never encounters the $ A_s-D_0(\lambda) $ degeneracy in the physically dangerous form identified here. The implied $ A_s $ diagnostic bridges this gap directly. Two limitations restrict the scope of present work. The parameter $ \varepsilon_{sc} $ is the exact $ f(Q) $ counterpart of the Song-Hu-Sawicki B-parameter as in \cite{Song:2006ej}. It inherits its status as a proximity measure rather than a stability criterion and does not replace a full perturbation analysis. It only serves as a qualitative tool in proceeding with cosmological analysis for the $ f(Q) $ class of models in light of the strong coupling issues found in \cite{Gomes:2023tur}. We note that the amplitude prior reduces $ \varepsilon_{sc}(0) $ from $ 0.120-0.188 $ to $ 0.028-0.042 $, indicating that in addition to resolving the degeneracy, the prior also moves the posterior away from the pathological $ |\varepsilon_{sc}| = 1 $ boundary. This is an observation of consistency rather than rigorous validity of the $ \ln(A_s) $ prior. Similarly, $ (aH/ck)^2\ll 1 $ ensures that we are well inside the deep sub-horizon limit, keeping in mind that the QSA is not well-defined in the present theory, as given in \cite{BeltranJimenez:2019tme}.

The Compressed CMB + $ \ln(A_s) $ prior does present a computationally inexpensive alternative to the full Boltzmann treatment, albeit with a small trade-off. \autoref{eq_asimplied} requires a reference clustering amplitude $ \sigma_{80}^{\mathrm{ref}} $ that can be tabulated once with any Boltzmann solver along with a growth integration per sample to get the total growth ratio $ R(\theta, \lambda) $. 
\begin{table*}
    \centering
    \renewcommand{\arraystretch}{1.25}
    \begin{tabular}{>{\centering\arraybackslash}m{12em}
    >{\centering\arraybackslash}m{8em}
    >{\centering\arraybackslash}m{8em}
    >{\centering\arraybackslash}m{8em}
    >{\centering\arraybackslash}m{8em}}
    \hline
        \textbf{Model} & $ \bm{q_0} $ & $ \bm{z_t} $ & $ \bm{\omega_{eff0}} $ & $ \bm{S_8} $ \\
        \hline 
        \multicolumn{5}{c}{\textbf{\textit{BD2R}}}\\
        \hline
        $ \bm{\Lambda} $\textbf{CDM} & $ -0.547 $ & $ 0.665 $ & $ -0.698 $ & $ 0.799 $ \\
        $ \bm{\Lambda} $\textbf{CDM}$ \bm{+\lambda_0} $ & $ -0.545 $ & $ 0.662 $ & $ -0.697 $ & $ 0.861 $ \\
        $ \bm{\Lambda} $\textbf{CDM}$ \bm{+\lambda_0+\ln(A_s)} $ & $ -0.545 $ & $ 0.662 $ & $ -0.697 $ & $ 0.800 $\\
        $ \bm{f(Q)} $ & $ -0.509 $ & $ 0.660 $ & $ -0.673 $ & $ 0.804 $\\
        $ \bm{f(Q)+\lambda_0} $ & $ -0.513 $ & $ 0.656 $ & $ -0.675 $ & $ 0.873 $\\
        $ \bm{f(Q)+\lambda_0+\ln(A_s)} $ & $ -0.523 $ & $ 0.656 $ & $ -0.682 $ & $ 0.800 $\\
        \hline
        \multicolumn{5}{c}{\textbf{\textit{BSDp}}}\\
        \hline
        $ \bm{\Lambda} $\textbf{CDM} & $ -0.531 $ & $ 0.638 $ & $ -0.687 $ & $ 0.873 $ \\
        $ \bm{\Lambda} $\textbf{CDM}$ \bm{+\lambda_0} $ & $ -0.527 $ & $ 0.631 $ & $ -0.684 $ & $ 0.924 $ \\
        $ \bm{\Lambda} $\textbf{CDM}$ \bm{+\lambda_0+\ln(A_s)} $ & $ -0.529 $ & $ 0.635 $ & $ -0.686 $ & $ 0.831 $\\
        $ \bm{f(Q)} $ & $ -0.487 $ & $ 0.626 $ & $ -0.658 $ & $ 0.886 $\\
        $ \bm{f(Q)+\lambda_0} $ & $ -0.480 $ & $ 0.621 $ & $ -0.654 $ & $ 0.943 $\\
        $ \bm{f(Q)+\lambda_0+\ln(A_s)} $ & $ -0.508 $ & $ 0.628 $ & $ -0.672 $ & $ 0.832 $\\
        \hline
    \end{tabular}
    \caption{Summary of present day best-fit cosmological parameters.}
    \label{table_presentdayvalues}
\end{table*}

\section{Conclusion}\label{sec_conclusion}
We have demonstrated that the background-inert $ \lambda_0\sqrt{QQ_0} $ extension to the coincident symmetric teleparallel $ f(Q) $ gravity introduces a systematic bias in parameter inference when using compressed CMB priors. The mechanism is the $ A_s-D_0(\lambda) $ degeneracy -- because the compressed CMB priors only constrain background geometry through the shift parameters $ R $ and $ l_a $, the primordial amplitude is left unconstrained, driving the sampler into exploiting the $ \sigma_{80}-\lambda_0 $ degeneracy valley. The effective gravitational coupling is suppressed at late-times and the sampler compensates by inflating $ \sigma_{80} $ while simultaneously preferring larger $ \lambda_0 $ values.

The degeneracy valley results in a genuine improvement to the fit. $ \chi^2_{min} $ is lowered by $ 2.5-4.2 $ for a single extra parameter. The comparison metrics, however, disagree on the preference over the concordance model. AIC and DIC mildly favor the $ +\lambda_0 $ variants in three out of four cases, BIC penalizes heavily across the board, while the Bayesian evidence disfavors them in all cases. None of the metrics actually record the cost for the improvement, which is precisely the gap that the $ \ln(A_s) $ prior fills. Specifically, this improvement costs a $ 32\%-63\% $ increase in the $ A_s^{\mathrm{implied}} $ as compared to $ A_s^{\mathrm{Planck}} $, exhibiting $ 1.5\sigma-2.6\sigma $ tension depending on the dataset and model combination.

The central methodological implication of this work is that the compressed CMB distance priors, while correctly applicable to background convergent theories by the corollary in \autoref{subsec_consistency_criterion}, are insufficient for theories in which a perturbative degree of freedom enters solely through $ \mu_G(z) $ decoupled from the background expansion history. In such cases, the primordial amplitude absorbs the modified growth without any penalty. Future analyses of this class of theories -- particularly as the next generation surveys deliver sub-percent growth measurements -- should either use the full shape CMB likelihood, or include the Planck prior on $ \ln(A_s) $ as a minimum safeguard. The implied $ A_s $ diagnostic provides an inexpensive post-processing check requiring only a tabulated reference amplitude and a growth integration, applicable to any existing posterior chain.

Finally, we wish to emphasize that this diagnostic is a phenomenological consistency test. The small values for $(aH/ck)^2$ and $|\varepsilon_{sc}|$ show that the posteriors stay well within the sub-horizon regime and away from the pathological $|\varepsilon_{sc}|\rightarrow1$ boundary. This does not, however, guarantee perturbative stability or directly quantify the strong coupling of non-linear $f(Q)$ models. Therefore, the various couplings considered here are just phenomenological parameterizations of a background-inert $\mu_G(z)$. Additionally, the Planck amplitude prior includes lensing of the CMB assuming a flat $\Lambda$CDM background interpreted with a growth of structure of GR, and so it is an approximation for these couplings.
\section*{Data Availability}
The observational datasets used in this article are publicly available. The Pantheon$ ^+ $ supernova compilation is available at \url{https://github.com/PantheonPlusSH0ES/DataRelease}. The BAO measurements from SDSS DR16 and DESI DR2 are available at \url{https://github.com/CobayaSampler/bao_data}, while the RSD compilation, the $ E_g $ statistic and compressed CMB data are all available in the respective cited papers.

The Bayesian inference was performed using the publicly available \texttt{COSMIX} package \cite{kolhatkar_cosmix_2026}, which can be accessed via GitHub (\url{https://github.com/AmeyaKolhatkar/COSMIX}) and Zenodo (\url{https://doi.org/10.5281/zenodo.19791571}). The Boltzmann solver \texttt{CAMB}, the dynamic nested sampling package \texttt{dynesty} and the visualization package \texttt{GetDist} are available in their respective public releases. 

\bibliographystyle{mnras}
\bibliography{ref}

\appendix 
\section{Robustness to the sound horizon}\label{app_rd}
The sound horizon enters only the BAO likelihoods. We test its role by importance reweighting each chain as
\begin{equation}
     w_i\rightarrow w_i\exp\left[\ln\mathcal{L}_{\rm BAO}(\theta_i,r_d')-\ln\mathcal{L}_{\rm BAO}(\theta_i,r_d)\right]
\end{equation}
where $ r_d $ is the value used in the chain and $ r_d' $ is the alternative. We consider two such alternatives: (i) a common $ r_d'=147.05 $\;Mpc for both SDSS DR16 and DESI DR2, and (ii) $ r_d' $ computed for each sample using the corresponding $ \omega_b, \omega_m $ values with \texttt{CAMB} \cite{Lewis:1999bs}, assuming a background with massless neutrinos and $ N_{\rm eff}=3.046 $. The derived values are interpolated on a grid in $ (\omega_b,\omega_m) $ that reproduces direct \texttt{CAMB} calls to $ 0.002 $\,Mpc. \autoref{table_rd} compiles the derived sound horizon, the effective sample size after reweighting relative to the original chain, the shifts of the posterior means, the distance of the implied amplitude from Planck and the evidence comparisons of \autoref{table_ICE}. The largest shifts are in the background parameters constrained by the BAO scale, at most $ 0.4\sigma $ in $ H_0 $ and $ 0.5\sigma $ in $ \Omega_{m0} $. $ \sigma_{80} $ moves by at most $ 0.1\sigma $, $ \lambda_0 $ by at most $ 0.2\sigma $, the distance of the implied amplitude from Planck by at most $ 0.18 $, and the evidence comparisons by at most $0.24$, comparable to their uncertainties. The conclusions of this work do not depend on the choice of $r_d$.
\begin{table*}
    \centering
    \small
    \setlength{\tabcolsep}{6pt}
    \renewcommand{\arraystretch}{1.3}
    \begin{tabular}{llcccccccc}
        \hline
        Data ($r_d$) & Model & $r_d^{\rm derived}$ [Mpc] & ESS & $\delta H_0$ & $\delta\Omega_{m0}$ & $\delta\sigma_{80}$ & $\delta\lambda_0$ & $\Delta N_\sigma$ & $\Delta\mathcal{B}$ \\
        \hline
        BSDp ($r_d=147.05$) & $\Lambda$CDM & --- & 0.92 & $+0.27$ & $-0.26$ & $+0.09$ & --- & $0.11$ & $0$ \\
         & $\Lambda$CDM$+\lambda_0$ & --- & 0.92 & $+0.27$ & $-0.26$ & $+0.07$ & $-0.02$ & $0.03$ & $-0.06$ \\
         & $\Lambda$CDM$+\lambda_0+\ln(A_s)$ & --- & 0.91 & $+0.30$ & $-0.28$ & $+0.08$ & $-0.11$ & $0.00$ & $-0.09^{\ddagger}$ \\
         & $f(Q)$ & --- & 0.91 & $+0.23$ & $-0.26$ & $+0.07$ & --- & $0.06$ & $-0.10$ \\
         & $f(Q)+\lambda_0$ & --- & 0.91 & $+0.21$ & $-0.24$ & $+0.05$ & $-0.03$ & $0.01$ & $-0.18$ \\
         & $f(Q)+\lambda_0+\ln(A_s)$ & --- & 0.90 & $+0.26$ & $-0.29$ & $+0.10$ & $-0.08$ & $0.01$ & $-0.01^{\ddagger}$ \\
        \hline
        BD2R ($r_d$ derived) & $\Lambda$CDM & $147.19\pm0.21$ & 0.53 & $+0.26$ & $-0.41$ & $+0.04$ & --- & $-0.07$ & $0$ \\
         & $\Lambda$CDM$+\lambda_0$ & $147.14\pm0.21$ & 0.48 & $+0.35$ & $-0.51$ & $-0.03$ & $-0.07$ & $-0.02$ & $-0.13$ \\
         & $\Lambda$CDM$+\lambda_0+\ln(A_s)$ & $147.15\pm0.21$ & 0.47 & $+0.31$ & $-0.47$ & $+0.00$ & $-0.09$ & $0.01$ & $0.05^{\ddagger}$ \\
         & $f(Q)$ & $147.23\pm0.23$ & 0.48 & $-0.21$ & $-0.00$ & $+0.06$ & --- & $-0.18$ & $0.24$ \\
         & $f(Q)+\lambda_0$ & $147.21\pm0.23$ & 0.44 & $-0.22$ & $-0.01$ & $+0.06$ & $+0.02$ & $0.08$ & $0.23$ \\
         & $f(Q)+\lambda_0+\ln(A_s)$ & $147.18\pm0.23$ & 0.44 & $-0.19$ & $-0.05$ & $-0.09$ & $-0.19$ & $0.02$ & $-0.14^{\ddagger}$ \\
        \hline
        BSDp ($r_d$ derived) & $\Lambda$CDM & $146.95\pm0.22$ & 0.88 & $+0.34$ & $-0.33$ & $+0.09$ & --- & $0.12$ & $0$ \\
         & $\Lambda$CDM$+\lambda_0$ & $146.89\pm0.22$ & 0.86 & $+0.38$ & $-0.37$ & $+0.07$ & $-0.03$ & $0.03$ & $-0.09$ \\
         & $\Lambda$CDM$+\lambda_0+\ln(A_s)$ & $146.94\pm0.22$ & 0.85 & $+0.39$ & $-0.39$ & $+0.08$ & $-0.12$ & $0.00$ & $-0.08^{\ddagger}$ \\
         & $f(Q)$ & $147.04\pm0.24$ & 0.88 & $+0.23$ & $-0.28$ & $+0.07$ & --- & $0.08$ & $-0.07$ \\
         & $f(Q)+\lambda_0$ & $147.00\pm0.23$ & 0.88 & $+0.22$ & $-0.28$ & $+0.05$ & $-0.04$ & $0.01$ & $-0.16$ \\
         & $f(Q)+\lambda_0+\ln(A_s)$ & $146.99\pm0.23$ & 0.86 & $+0.28$ & $-0.34$ & $+0.09$ & $-0.10$ & $0.01$ & $-0.02^{\ddagger}$ \\
        \hline
    \end{tabular}
    \caption{Robustness of the results to the sound horizon. The published chains, with $r_d$ fixed at $147.05$\,Mpc (DESI DR2) and $147.8$\,Mpc (SDSS DR16), are importance-reweighted to a common $r_d=147.05$\,Mpc and to $r_d$ computed with \texttt{CAMB} for each sample from $(\omega_b,\omega_m)$ (massless neutrinos, $N_{\rm eff}=3.046$). $r_d^{\rm derived}$: posterior mean and standard deviation of the derived sound horizon. ESS: effective sample size relative to the original chain. $\delta$: shift of the posterior mean in units of the original standard deviation. $\Delta N_\sigma$: distance of the implied amplitude from Planck (\autoref{subsec_results2}), new minus old. $\Delta\mathcal{B}$: Bayes factor against $\Lambda$CDM on the same data, new minus old; $^{\ddagger}$log predictive density of the Planck amplitude instead (\autoref{table_ICE}).}
    \label{table_rd}
\end{table*}

\bsp
\label{lastpage}
\end{document}